\documentclass{aa}
\usepackage{graphicx}
\usepackage{amsmath}	
\usepackage{amssymb}	
\usepackage{relsize}
\usepackage{txfonts}
\usepackage[colorlinks=true,linkcolor=blue,allcolors=blue]{hyperref}%

\newcommand{\degree}{^\circ}

\newcommand{\Msun}{M$_\odot$}

\def\nph{n_{\rm ph}}
\def\nsp{n_{\rm sp}}
\def\nuph{\nu_{\rm ph}}
\def\nusp{\nu_{\rm sp}}
\begin{document} 

   \title{A vertically orientated dark matter halo marks a flip of the Galactic disc}
   \subtitle{}

   \author{Ling Zhu$^1$\thanks{lzhu@shao.ac.cn}, Runsheng Cai$^1$, Xi Kang$^2$\thanks{kangxi@zju.edu.cn}, Xiang-Xiang Xue$^{3,6}$\thanks{xuexx@nao.cas.cn}, Chengqun Yang$^4$, \\Lan Zhang$^3$, Shude Mao$^5$, Chao Liu$^3$ }

   \institute{
$^{1}$ Shanghai Astronomical Observatory, Chinese Academy of Sciences, 80 Nandan Road, Shanghai 200030, China \\
$^{2}$ Institute for Astronomy the School of Physics, Zhejiang University, 38 Zheda Road, Hangzhou, 310027, Zhejiang, China \\
$^{3}$ National Astronomical Observatories, Chinese Academy of Sciences, Beijing 100101, China\\
$^{4}$ School of Physics and Optoelectronic Engineering, Hainan University, 58 Renmin Avenue, Haikou, 570228, China\\
$^{5}$ Department of Astronomy, Westlake University, Hangzhou, Zhejiang 310030, China\\
$^{6}$ Institute for Frontiers in Astronomy and Astrophysics, Beijing Normal University, Beijing 102206, China
}

   \date{Received; accepted}
   
   \titlerunning{A Vertically Orientated Dark Matter Halo}
\authorrunning{Zhu et al.}  
 
  \abstract
   {Unveiling the 3D shape of the Milky Way's dark-matter halo is critical to understanding its formation history. We created an innovative dynamical model that makes minimal assumptions about the internal dynamical structures and  accommodates a highly flexible triaxial DM halo. By applying the method to 6D phase-space data of K-giant stars from LAMOST + Gaia, we robustly determined the 3D dark-matter distribution of the Milky Way out to approximately $50$ kpc. We discovered a triaxial, nearly oblate dark-matter halo with $q_{\rm DM} = Z/X= 0.92\pm0.08$, $p_{\rm DM} = Y/X= 0.8\pm0.2$ on average within 50 kpc, where the $Z$ axis is defined perpendicular to the stellar disc. The axes ratio $q_{\rm DM} > p_{\rm DM}$ is strongly preferred; the long-intermediate axis plane of the dark-matter halo is unexpectedly vertical to the Galactic disc, yet aligned with the `plane of satellites'. This striking configuration suggests that the Galactic disc (and the inner halo) has flipped, likely torqued by minor mergers, from an original alignment with the outer dark-matter halo and satellite plane, as is supported by Milky Way analogues from Auriga and TNG50. By allowing $q_{\rm DM}(r)$ and $p_{\rm DM}(r)$ to vary with radii, we find tentative evidence that the dark-matter halo is twisted. This agrees alignment with the disc in the inner regions and transitions to a vertical orientation at $r\gtrsim 20$ kpc, supporting the disc flip scenario prediction. Such disc reorientation is non-trivial, yet its physical mechanism is straightforward to comprehend and naturally originates a vertical satellite plane. Our findings offer a unified framework that links dark-matter halo orientation, satellite alignment, and disc evolution, reinforcing the internal consistency of the Milky Way in the $\Lambda$ cold dark matter model.
   }


   \maketitle


\section{Introduction}

The 3D shape of the dark-matter (DM) halo serves as a fundamental test of the $\Lambda$ cold dark matter ($\Lambda$CDM) model \citep{Peter2013MNRAS.430..105P} and retains key information about a galaxy's filamentary accretion history from the cosmic web and interactions with massive satellites \citep{Arora2025arXiv250420133A}.
 Functioning as an intermediary between the Galaxy and its satellites, the shape of the DM halo also plays a key role in deciphering the long-standing conundrum of a vertically aligned `plane of satellites' with the Galactic disc, in the case of the Milky Way \citep{Shao2016MNRAS.460.3772S}. 

Determining the exact shape of the DM halo from observations, even for the Milky Way, is a challenge. One major method of probing the DM distribution is to model the motions of tracers, which reflect the underlying gravitational potential. Stellar streams, viewed as simple orbital pathways that trace the halo, played a significant role in constraining the shape of the Milky Way DM halo \citep{Bovy2016ApJ...833...31B, Bowden2015MNRAS.449.1391B, Koposov2010ApJ...712..260K, Palau2023MNRAS.524.2124P, Law2010ApJ...714..229L}. However, the most commonly used dynamically cold streams, for example Pal 5 and GD--1, trace only the inner halo at $r\lesssim 20$ kpc. And a single stream is easily disturbed \citep{Nibauer2024ApJ...969...55N} and may lead to biassed results \citep{Vasiliev2021MNRAS.501.2279V}.

Halo stars -- with the full 6D phase-space information observed -- span the entirety of space and, statistically, should exhibit strong power in constraining the underlying 3D DM distribution. However, these stars exist with complex orbital dynamics, necessitating accurate dynamical modelling. 
At present, the Milky Way halo is analysed using two primary types of dynamical models: the Jeans model and the distribution function (DF)-based dynamical model. These models yield divergent results regarding the shape of the DM halo, ranging from oblate with $q \sim 0.4-0.7$ \citep{Loebman2014ApJ...794..151L, Hattori2021MNRAS.508.5468H, Li2022MNRAS.510.4706L}, to nearly spherical with $q \sim 0.9-1$ \citep{Wegg2019MNRAS.485.3296W, Zhang2025}, and to prolate with $q \sim 1.3$ \citep{Posti2019A&A...621A..56P}. All these models are restricted by axisymmetric assumptions and incorporate ad hoc assumptions about the intrinsic dynamical structures. Despite extensive research over the years, the true shape of the Milky Way DM halo remains debated \citep{Hunt2025NewAR.10001721H}. 

Due to the large uncertainty surrounding the DM halo shape of the Milky Way from observations, its link to the spatial arrangement of satellite galaxies has not been seriously considered. The satellite galaxies in the Milky Way are not distributed isotropically but rather form a thin plane vertical to the Galactic disc \citep{Lynden-Bell1976MNRAS.174..695L,Kroupa2005A&A...431..517K,Pawlowski2018MPLA...3330004P, Sawala2023NatAs...7..481S}. The thickness of this plane of satellites has been extensively studied; it was previously thought to be rare but has now become more frequently observed through cosmological simulations \citep{Sawala2023NatAs...7..481S}. The cause of its perpendicular alignment to the Galactic disc remains uncertain \citep{Vasiliev2021MNRAS.501.2279V}. 
Typically, both the stellar disc and satellite system align parallel to the DM halo, sharing a similar net angular momentum direction, leading to a co-alignment of the stellar disc and satellite plane \citep{Shao2016MNRAS.460.3772S}. However, in systems resembling the Milky Way, in which the satellite system lies perpendicular to the stellar disc, the DM halo might exhibit a twist, being co-aligned with the stellar disc in the inner regions but aligned vertically in the outer regions \citep{Shao2021MNRAS.504.6033S}. Precisely measuring the DM shape as a function of radius could provide essential insights into the origin of the vertical satellite plane and elucidate its relationship with the large-scale matter distribution. Additionally, this suggests that the DM halo of the Milky Way may be triaxial, with a shape that changes from the inner to outer regions, necessitating dynamic models with more flexible DM halos.

 Recently, a triaxial Schwarzwald model that numerically represents the DF of tracers by superposition of stellar orbits has been applied to the Milky Way halo, indicating a triaxial, nearly prolate halo with a large tilt angle \citep{Dillamore2025arXiv251000095D}. Nevertheless, there exists a strong degeneracy between the axis ratios of the DM halo and the tilt angle. This model was mainly constrained by the shape of a tilted stellar halo \citep{Han2022AJ....164..249H} associated with the Gaia-Sausage-Enceladus (GSE) \citep{Helmi2018Natur.563...85H, Belokurov2018MNRAS.478..611B}.

In a previous paper \citep{Zhu2025A&A...703A..43Z}, we presented and validated an empirical triaxial orbit-superposition model tailored for the Milky Way halo observed by the Large Sky Area Multi-Object Fibre Spectroscopic Telescope (LAMOST) and the Global Astrometric Interferometer for Astrophysics (Gaia). This approach accommodates a highly flexible triaxial DM halo, where the axis ratios are allowed to vary with radius, and it uses the full 6D phase-space information from the observations.
Compared to traditional made-to-measure and Schwarzschild methods, this method does not theoretically sample the orbit libraries or determine the orbit weights by fitting the model to data; instead, the orbit library and orbit weights are entirely determined from the observations. The results are thus highly data-driven.

In this paper, we apply the empirical triaxial orbit-superposition model to the Milky Way halo, utilising 6D phase-space data from LAMOST and Gaia, and robustly determine the 3D DM distribution of the Milky Way, extending to approximately 50 kpc. We then compare the findings with Milky Way analogues from the Auriga and TNG50 cosmological simulations and establish a coherent scenario linking the DM halo shape and satellite arrangement. The paper is organised in the following way: we describe the details of observational data in Section 2, introduce model construction in Section 3, and show the results in Section 4. We compare the 3D shape of the DM halo and the arrangement of the Milky Way's satellite system with cosmological simulations in Section 5. We discuss in Section 6 and conclude in Section 7.

\section{Observational data}

\subsection{Data preparation}
\begin{figure}
\centering\includegraphics[width=8cm]{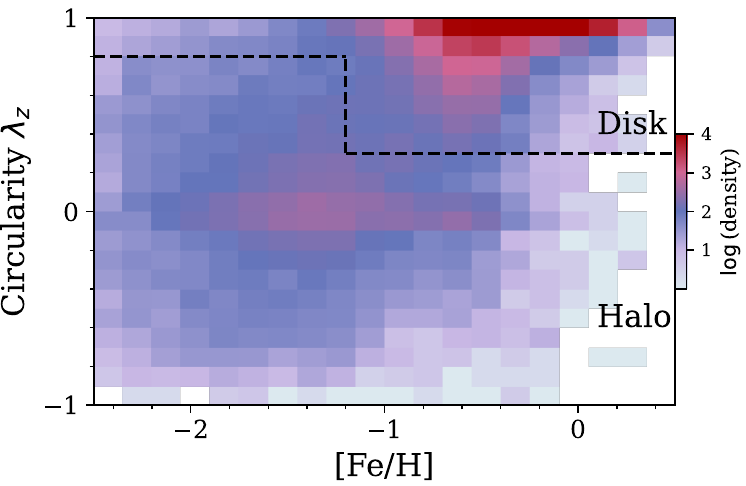}
\caption{Density of stars in the parameter phase of circularity, $\lambda_z$, versus metallicity, [Fe/H]. We first removed the disc stars with a circularity of $\lambda_z > 0.8$, then also removed the group of stars with  $\lambda_z>0.3$ and $[\rm Fe/H]>-1.2$. The rest were kept as halo stars in our sample.
}
\label{fig:cutdisk}
\end{figure}

\begin{figure}
\centering\includegraphics[width=9cm]{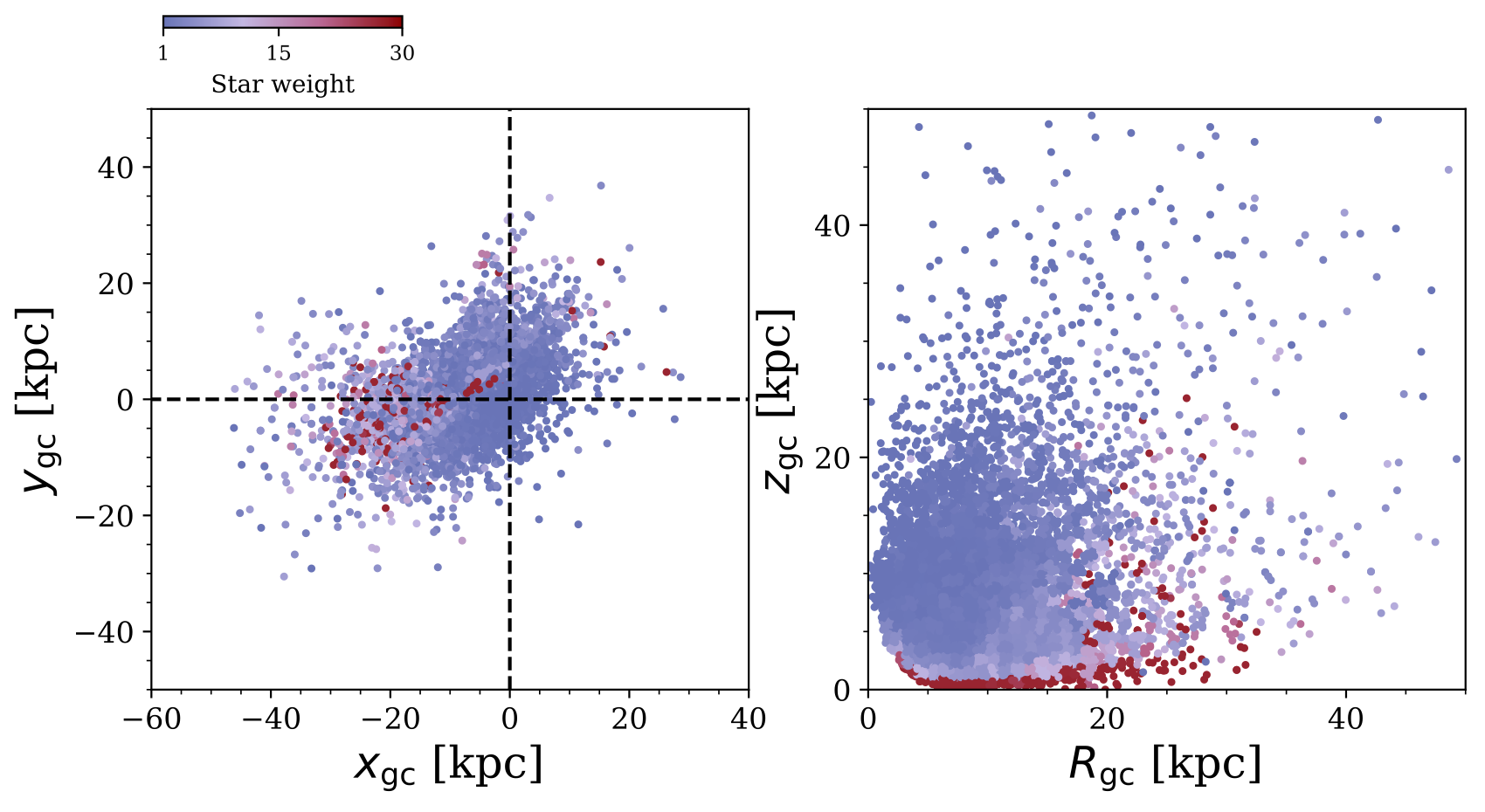}
\caption{K giants in our final smooth halo sample in the northern hemisphere, coloured by the weights of particles obtained by correction of the selection function. The sample can be taken as complete in the $R_{\rm gc}-z_{\rm gc}$ space after correction of the selection function. 
}
\label{fig:sample}
\end{figure}

We obtained the 3D positions and 3D velocities of halo stars by combining observations from LAMOST and Gaia.
LAMOST has gathered spectroscopic data for approximately 10 million stars within the Milky Way. We used k giants from LAMOST DR9; the distance is accurately measured with a relative uncertainty of $\sim 15\%$ \citep{Zhang2023}. By cross-matching with Gaia DR3, we obtained a sample of 619,284 k-giants with 3D velocity measurements; the typical errors in the radial velocity, $v_r$, tangential velocity, $v_{\phi}$, and $v_{\theta}$ are 10, 30, and 30 km/s, respectively. 
 We excluded disc stars chemodynamically to avoid biassing the spatial distribution of the halo. We first removed all of the stars in the cold disc with $\lambda_z > 0.8$, then removed the group of stars with $\lambda_z>0.3$ and $[\rm Fe/H]>-1.2$ that are mostly within 20 kpc, not in dynamical equilibrium, and likely to be induced from the disc (see Fig.~\ref{fig:cutdisk}).
The remaining stars were considered halo stars. We carefully excluded all substructures, including streams and overdensities, using the friend-of-friend method \citep{Sun2025ApJ...979..213S, Wang2022MNRAS.513.1958W, Yang2019ApJ...880...65Y}. We further removed stars with velocity errors of $>100$ km/s and performed $3\sigma$ clipping to remove the outliers. We preserved a refined halo sample comprising 14,497 K-giants, with 11,415 in the northern hemisphere and 3,082 in the southern hemisphere. The sample spans a vast area in the northern hemisphere, up to $r\lesssim 50$ kpc.

\subsection{Correction of the selection function}
We corrected for the selection effects of the LAMOST survey following previous studies \citep{Liu2017RAA....17...96L}, in which the photometric data were assumed to be complete within their limiting magnitude. In the colour-magnitude plane $(c, m)$ along each line of sight $(l,b)$, the photometric density, $\nuph$, is equal to the spectroscopic density, $\nusp$, times the inverse of the selection function, $S^{-1}$,
\begin{equation}
    \nuph(d|c, m, l, b) = \nusp(d|c, m, l, b) S^{-1}(c, m, l, b).
\end{equation}

The spectroscopic density, $\nusp(d|c, m, l, b)$, was calculated using the kernel density estimation (KDE) method:
\begin{equation}
    \nusp(d|c, m, l, b) = \frac{1}{\Omega d^2} \sum_{i}^{\nusp(c,m,l,b)} p_i(d),
\end{equation}
where $p_i(d)$ is the probability density function of distance for the $i_{\rm th}$ star, and $\Omega d^2$ is the volume element between $d$ and $d + \Delta d$.

The selection function, $S^{-1}(c, m, l, b)$, was evaluated from
\begin{equation}
    S^{-1}(c, m, l, b) = \frac{\nsp (c,m,l,b)}{\nph (c,m,l,b)},
\end{equation}
where $\nsp (c,m,l,b)$ and $\nph (c,m,l,b)$ are the numbers of stars with spectroscopic and photometric data, respectively, within each bin of $(c,m,l,b)$.

The stellar density along a given line of sight is thus
\begin{equation}
    \nuph(d|l, b) = \iint \nusp(d|c, m, l, b) S^{-1}(c, m, l, b){\rm d}c{\rm d}m.
\end{equation}
The selection-corrected average density distribution of the stellar halo in the $(R_{\rm gc}, z_{\rm gc})$ plane ($\bar \nuph (R_{\rm gc}, z_{\rm gc}) $) could then be constructed \citep{Xu2018MNRAS.473.1244X, Yang2022AJ....164..241Y}. 

We binned the stars observed spectroscopically in the $(R_{\rm gc}, z_{\rm gc})$ plane and obtained the number of stars in each bin $\nsp (R_{\rm gc}, z_{\rm gc})$. For each of the stars in the bin, $i$, we calculated their photometric weight:
\begin{equation}
    \omega_i = \frac{\bar \nuph (R_i, z_i) \times V}{\nsp(R_i, z_i)}, 
\label{eq:weight}
\end{equation}
where $V= 2\pi R_i dRdz$ is the volume of the bin. We considered all the stars in a $1.0 \times 1.0$ kpc bin to have the same weight.

We present our sample of k-giants in the northern hemisphere in Fig.\ref{fig:sample}. The observations close to the disc plane and at $r\gtrsim 30$ kpc are less complete; thus, the stars have higher weights. With weights correcting for their bias from observational selection, the sample in the northern hemisphere can be considered representative of the halo stars in 6D position-velocity distributions.

\subsection{Correction of the LMC bias}
The influence of LMC on the stellar halo at $r<50$ kpc is small. We binned the data in the $R_{\rm gc}-z_{\rm gc}$ plane with a bin size of $10\times 10$ kpc and evaluated their mean velocities, $v_x$, $v_y$, and $v_z$, within each bin. We observe a minor net velocity in $v_z$ and $v_x$, around $\sim 10-20$ km/s, whereas a system in perfect dynamical equilibrium should have these velocities at 0. These offsets are presumably caused by the LMC's influence \citep{Erkal2020MNRAS.498.5574E}. We adjusted for this by deducting the mean $v_z$ and $v_x$ velocities from each star's velocity, ensuring zero mean velocities in every bin. 
Although the LMC's effect is limited in the areas under examination, correcting for it does not significantly alter our final results in constraining the 3D distribution of the DM halo. However, the agreement between the velocity distributions of the model and the data improves slightly when this correction is applied, as was expected.

\section{Model construction}
Our sample serves as a good approximation of being stationary, i.e. the stars' DF $f(\boldsymbol x,\boldsymbol v)$ will statistically not change when orbiting them in the correct gravitational potential. Driven by data insights, we presented a novel dynamical model \citep{Zhu2025A&A...703A..43Z} aimed at constraining the gravitational potential by directly comparing the observed data distribution, $f_{\rm data}(\boldsymbol x,\boldsymbol v)$, with the distribution derived from a model superposed by the orbits of these stars integrated within the potential $f_{\rm model}(\boldsymbol x,\boldsymbol v | \rm potential)$. This approach, named the empirical triaxial orbit superposition method -- based on minimal assumptions of dynamical models that the system is stationary -- overcomes key limitations of previous models. The method has been validated and proves effective at uncovering the 3D DM distribution in galaxies similar to the Milky Way. We have applied it to mock data created from Auriga simulations, selecting three galaxies with different 3D shapes of the underlying DM halo. In all three cases, our model accurately recovers the 3D DM distribution, including the axis ratios and the radial density profile. For one case with a twisted DM halo from the inner to outer regions, our model succeeds in capturing the radial variation in the DM halo shape. 
 
\subsection{Empirical triaxial orbit superposition method}

Here we briefly introduce the main steps to create the empirical orbit superposition model: (1) we built a model for the gravitational potential with a few free parameters; (2) given one set of parameters, we used the 6D phase space information of stars in the northern hemisphere as initial conditions, and calculated their stellar orbits in the gravitational potential; (3) we superposed the stellar orbits with their weights given by the selection function correction and constructed stellar density distributions and velocity distributions from the orbit superposition model; (4) we compared the data and model to calculate the likelihood or $\chi^2$ of each model; (5) we explored the parameter grid of the gravitational potential and found the best-fitting models with the maximum likelihood or least $\chi^2$. 

We constructed a model of the gravitational potential by including a barred bulge, a thin disc, a thick disc, and a triaxial DM halo with the following density profiles:
\begin{equation}
\rho_{\rm bulge} = \frac{105M}{32\pi pq a^3}[1- (\frac{\tilde{r}}{a})^2]^2;
\end{equation}

\begin{equation}
\rho_{\rm disc}= \frac{\Sigma_0 }{2h} \times \exp(-[\frac{R}{R_d}]^{1/n} - \frac{R_{\rm cut}}{R}) \times \exp(-\frac{|z|}{h}).
\end{equation}

We fixed the density distribution of the bulge and discs obtained from previous studies \citep{Bland-Hawthorn2016}: with $M_{*,\rm bulge} = 1.6 \times 10^{10} M_{\odot}$, a scale radius of $a = 3.5$ kpc, axis ratios of $p=0.44$, $q=0.31$ for the bulge, $M_{*,\rm thin} = 3.16 \times 10^{10} M_{\odot}$, a scale radius of $R_d = 2.6$ kpc, a scale height of $h = 0.3$ kpc, an inner cut-off radius of $R_{\rm cut} = 7$ kpc \citep{Lian2024NatAs...8.1302L}, a Sersic index of $n=1$ for the thin disc and 
$M_{*,\rm thick} = 6\times10^9 M_{\odot}$, a scale radius of $R_d = 2.0$ kpc, a scale height of $h = 0.9$ kpc, and a Sersic index of $n=1$ for the thick disc\footnote{We tried alternative models, including a model combining a disc without an inner cut-off and a Sersic bulge, as well as a model with circular velocity at the solar position ($r=8.2$ kpc \citep{Bland-Hawthorn2016}) fixed at $V_c =235$ km/s \citep{McMillan2017MNRAS.465...76M}; none of them makes noticeable differences to our final results for the DM shape. }.

We adopted a flexible triaxial generalised NFW model for the DM halo, allowing free orientations of the DM halo and a variable 3D shape as a function of radius, generally following \citep{Vasiliev2021MNRAS.501.2279V}:
\begin{equation}
\rho_{\rm halo} = \rho_0 (\frac{\tilde{r}}{r_s})^{-\gamma} [1+(\frac{\tilde{r}}{r_s})^{\alpha}]^{\frac{\gamma-\beta}{\alpha}} \times \exp{-(\frac{\tilde{r}}{r_{\rm cut}})^{\xi}},
\end{equation}
in which
$\tilde{r} =\sqrt{X^2 + (Y/p_{\rm DM})^2 + (Z/q_{\rm DM})^2}$, $p_{\rm DM}$, and $q_{\rm DM}$ are the ratios of the axes. 
The co-ordinates $\boldsymbol X\equiv \{X,Y,Z\}$ are related to the usual Galactocentric Cartesian co-ordinates $\boldsymbol x_{\rm gc}\equiv \{x_{\rm gc},y_{\rm gc},z_{\rm gc}\}$ by a rotation matrix, $\boldsymbol X = \mathsf R \boldsymbol x_{\rm gc}$, parameterised by three Euler angles, $\alpha_q$, $\beta_q$, and $\gamma_q$.
The $Z$ axis is tilted by the angle $\beta_q$ relative to the $z_{\rm gc}$ axis. Meanwhile, the angles $\alpha_q$ ($\gamma_q$) define the rotation between the $x_{\rm gc} (X) $ axis and the intersection line of the $x_{\rm gc}y_{\rm gc}$ and $XY$ planes.

The tilt angle $\beta_q$ highly degenerates with the vertical axis ratio, $q_{\rm DM}$. We fixed $\beta_q = 0\degree$ and allowed $q_{\rm DM}>1$ in our model. The $Z$ axis is thus always defined as perpendicular to the stellar disc. Because the direct observations are not complete and we have averaged the information in the azimuthal direction, our model cannot constrain the azimuthal orientation of the DM halo. We fixed $\alpha_q=0\degree$ in our fiducial model. In this model with fixed $\beta_q$ and $\alpha_q$, the axis ratio $q_{\rm DM}$ and $p_{\rm DM}$ will reflect the DM halo orientation. 

For the radial profile, we fixed $\alpha=1$, $\beta=3$, and chose the outer cut-off radius $r_{\rm cut}=500$ kpc and the cut-off strength $\xi=5$. We were left with five free parameters in the DM halo: $\rho_0$, $r_s$, $\gamma$, $p_{\rm DM}$, and $q_{\rm DM}$.

When needed, we allowed $p_{\rm DM}$ and $q_{\rm DM}$ to vary as a function of radius, with
\begin{eqnarray}
\label{eqn:pr}
p_{\rm DM}(r) = (p_{\rm in} + p_{\rm out} (\frac{r-10}{r_q})^2) / (1 + (\frac{r-10}{r_q} )^2)\\
p_{\rm DM}(r<10) = p_{\rm in}\\
q_{\rm DM}(r) = (q_{\rm in} + q_{\rm out} (\frac{r-10}{r_q})^2) / (1 + (\frac{r-10}{r_q} )^2)\\
q_{\rm DM}(r<10) = q_{\rm in}
\label{eqn:qr}
,\end{eqnarray}
where $r = \sqrt{X^2 + Y^2 + Z^2}$, $p_{\rm in}$, $p_{\rm out}$, $q_{\rm in}$, $q_{\rm out}$, and the scale radius, $r_q$, were allowed to be free parameters.

Given each set of parameters in the gravitational potential, we integrated the stellar orbits using the publicly released AGAMA package \citep{Vasiliev2019MNRAS.482.1525V}\footnote{https://github.com/GalacticDynamics-Oxford/Agama} for the 11415 stars in the northern hemisphere.
We integrated ten orbital periods for each star and withdrew 1000 particles from each orbit with equal time steps, and each particle was given the weight of the star that initialises the orbit. We superposed particles drawn from the orbits together, obtaining an orbit superposed model that numerically represents the DF in 6D phase space of the Milky Way stellar halo. 

\subsection{Evaluation of the goodness of the model}
\begin{figure}
\centering\includegraphics[width=7.5cm]{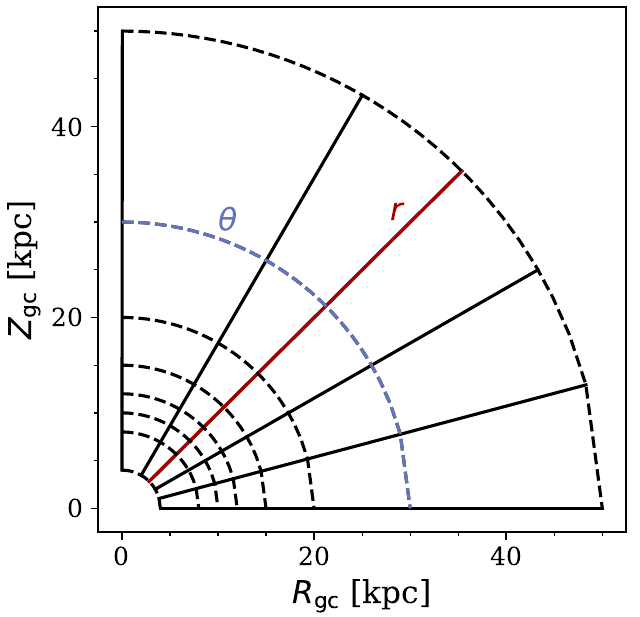}
\caption{Spatial binning scheme we used for comparing the velocity distributions of the model and data. We divided the model into $7\times 5$ spatial bins along $r\times \theta$, and calculated the likelihood of stars located within each bin by comparing to the velocity distributions of the model in the corresponding bin.
}
\label{fig:divide}
\end{figure}

\begin{figure*}
\centering\includegraphics[width=16cm]{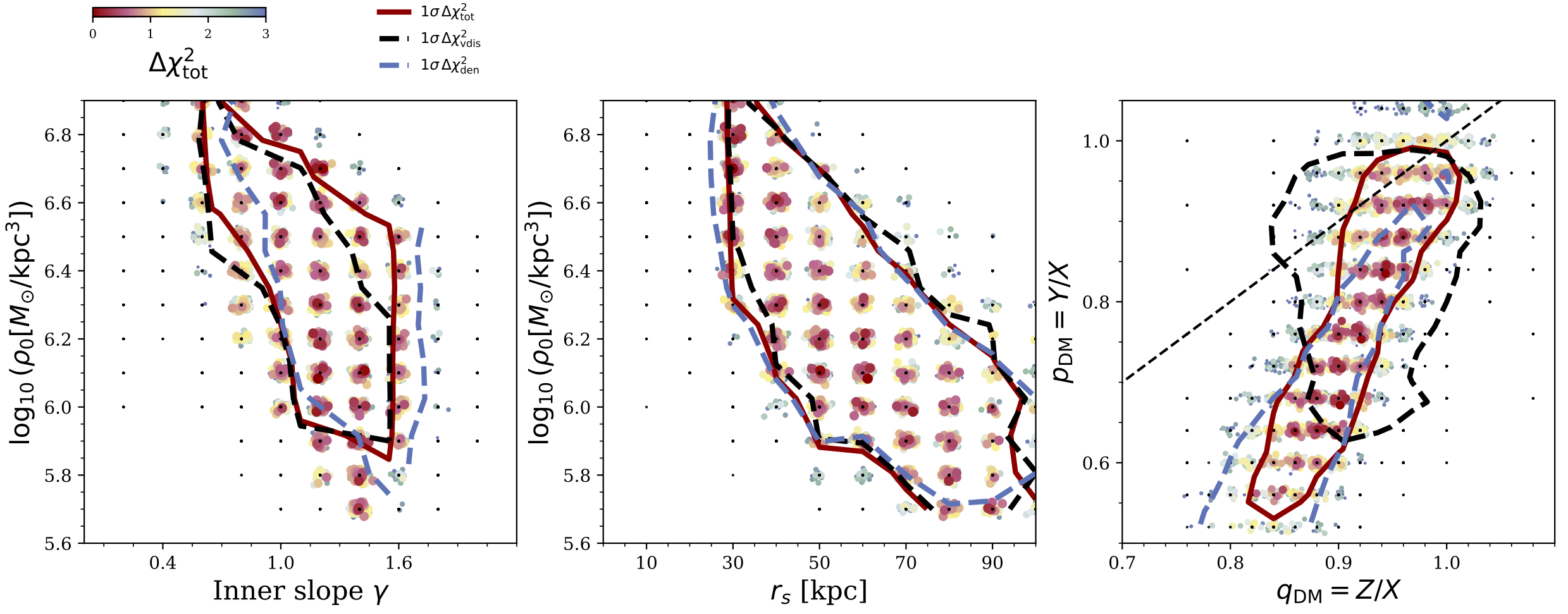}
\caption{Model constraints on the parameters of the underlying DM distribution, $\rho_0$, $r_s$, $\gamma$, $q_{\rm DM}$, and $p_{\rm DM}$. The three parameters determining the radial distribution, $\rho_0$, $r_s$, and $\gamma$, are degenerate due to the lack of data points at $r<8$ kpc and $r>50$ kpc. However the 3D DM distribution within our data coverage ($r\lesssim 50$ kpc) are well constrained by our model, with a significant role played by the density distribution, $\chi^2_{\rm den}$.
}
\label{fig:grid}
\end{figure*}

\begin{figure}
\centering\includegraphics[width=9cm]{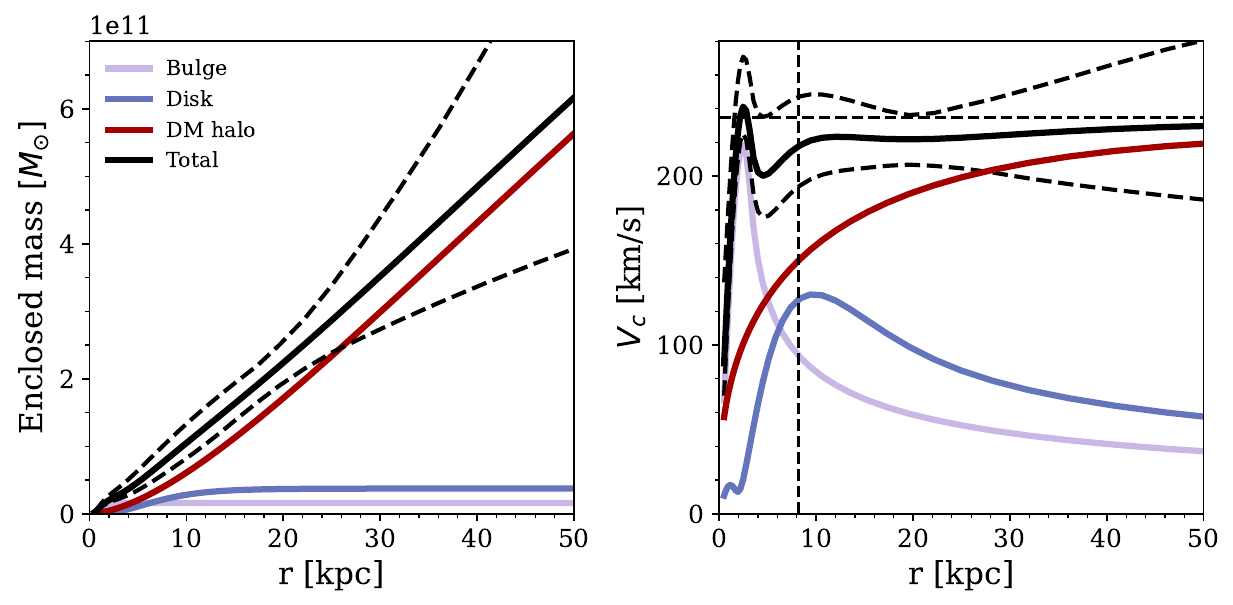}
\caption{Enclosed mass profiles and rotation curve that we obtained for the Milky Way. The solid line is the average of models within the $1\sigma$ confidence level, while the dashed curves are the upper and lower boundary.}
\label{fig:mass}
\end{figure}

We detail particular components of the DF: the spatial density distribution and velocity distributions at different positions, and we evaluate these for the model against the data. We calculated the stellar density distribution in the $R_{\rm gc}-z_{\rm gc}$ plane, dividing it into $25\times25$ bins within 50 kpc. The density in each bin was calculated by $N/(2\pi R dR dz)$ in units of $N/{\rm kpc}^3$. We calculated the density directly for both the observed data, $\rho_{\rm data}$, and the model, $\rho_{\rm model}$. The uncertainty of the data, $d\rho_{\rm data}$, was assessed by bootstrapping the distances of stars along with their uncertainties. We calculated $\chi^2_{\rm den}$ by directly comparing the data and the model:
\begin{equation}
\chi^2_{\rm den} = \sum (\rho_{\rm data} - \rho_{\rm model})^2/ d\rho_{\rm data}^2.
\end{equation}

To analyse velocity distributions, we categorised model particles into $N_{\rm bin} = 7\times 5$ bins in the direction of $r$ and $\theta$ as is shown in Fig.~\ref{fig:divide}. We constructed the velocity distributions ($v_r$, $v_{\phi}$, $v_{\theta}$) within each bin, $j$, and represent the distribution of each velocity component by a histogram using intervals of $M$, labelled ($v^k_{m,j}$, $h^k_{m,j}$), where $k$ corresponds to the velocities $v_r$, $v_{\phi}$, and $v_{\theta}$.

In bin $j$, we have $N_j$ stars, and for a star, $i$, with a velocity and measurement uncertainty ($v^k_i$, $\sigma^k_i$) within that bin, we determined its likelihood compared to the model for each of the three velocity components as follows:
\begin{equation}
P^k_{i, j} = \frac{1}{\sqrt{2\pi} \times \sigma^k_i} \times \frac{\sum_{m=1}^M h^k_{m,j} \times  \exp[- (v^k_i - v^k_{m,j})^2 / (2(\sigma^k_i)^2) ]} {\sum_{m=1}^M(h^k_{m,j})}
.\end{equation}
The collective likelihood of all $N_j$ stars in this bin is given by $L_j^k = \sum_{i=1}^{N_{\rm j}} \log(P^k_{i,j})$, and the likelihood of combining all bins is represented by $L^k = \sum_{j=6}^{N_{\rm bin}} L_j^k$. 

Our data for integrating stellar orbits primarily cover the range $4<r<50$ kpc, meaning that we lack data for orbits starting at $r<4$ kpc. These orbits could significantly influence the DF of stars within $r<8$ kpc. Consequently, the estimated DF from the model remains incomplete for $r<8$ kpc, whereas it is more than $90\%$ complete for the range $8<r<50$ kpc. Thus, we excluded the five bins in $r=[4,8]$ kpc from our likelihood calculations. Although our observations for the southern hemisphere are incomplete, the 3082 stars available should conform to the same DF. We incorporated these stars into the northern spatial bins and included them in our likelihood calculations. 

The likelihood across the three velocity components was merged to derive the constraints from the velocity distributions:
\begin{equation}
 \chi^2_{\rm vdis} = -2(L^{v_r} + L^{v_{\phi}} + L^{v_{\theta}}).
\end{equation}
Finally, we combined the $\chi^2$ values from both the density distribution and the velocity distributions to obtain
\begin{equation}
 \chi^2_{\rm tot} = -2(L^{v_r} + L^{v_{\phi}} + L^{v_{\theta}}) + \chi^2_{\rm den}.
\end{equation}

\section{Model results}
\subsection{The 3D dark matter distribution of Milky Way}

\begin{figure*}
\centering\includegraphics[width=18cm]{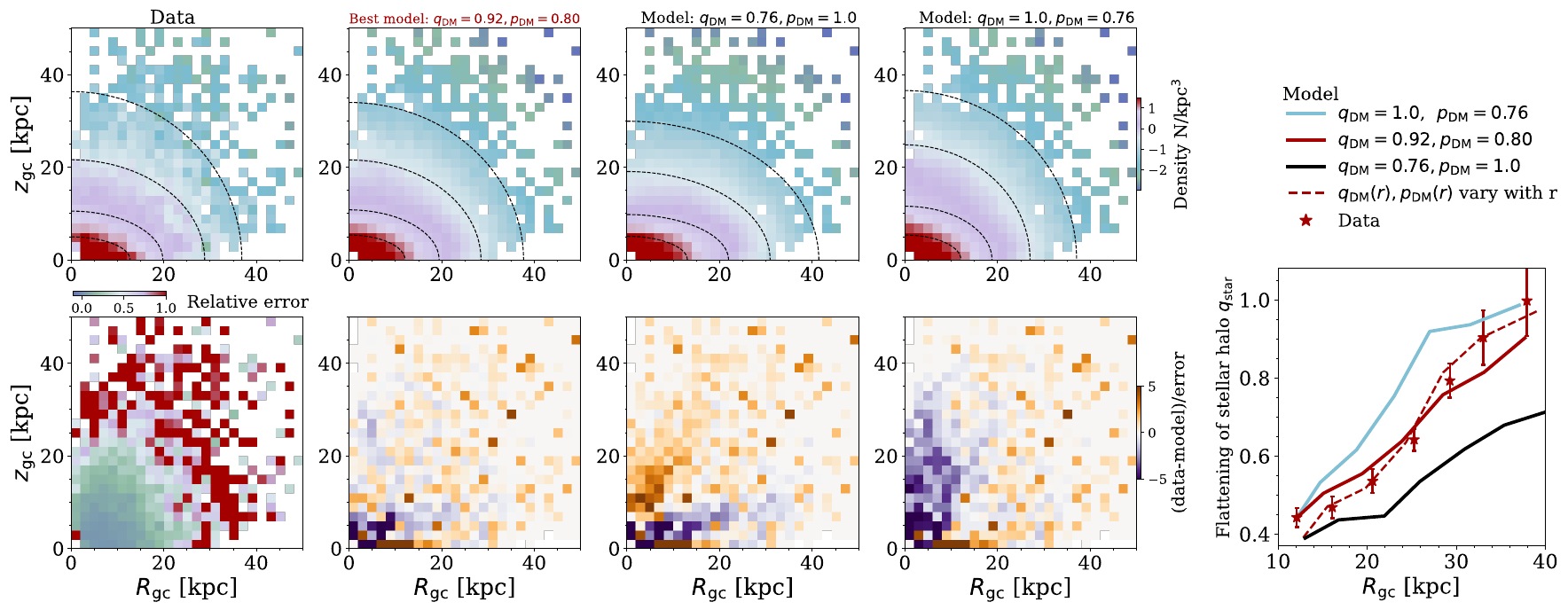}
\caption{\textbf{Comparison of the observational data and the empirical orbit superposed model: the stellar density distribution.} The top panels from left to right are the 2D density distribution $\rho_{\rm star}$ in $R_{\rm gc}$ versus $z_{\rm gc}$ constructed by data, our best-fitting model ($q_{\rm DM} = 0.92$, $p_{\rm DM} = 0.80$), a model with a flatter DM halo, and a model with a rounder DM halo. The bottom panels show the relative error of the data, and the corresponding model residuals. 
The inset panel in the right shows the flattening of the stellar halo, $q_{\rm star} \equiv z_{\rm gc}/R_{\rm gc}$, as a function of the radius, $R_{\rm gc}$, extracted from the iso-density contours of the 2D density distribution. The star symbols with the error bar are data. The solid red, light blue, and black curves indicate the three models shown on the left. The dashed red curve indicates a model allowing $q_{\rm DM}$ and $p_{\rm DM}$ to vary as a function of radius. The best-fitting model constructed stellar density distribution that matches the observational data well, while models with flatter or rounder DM halos constructed stellar halos that are either too flat or too round. The density distribution of the halo stars, especially the flattening $q_{\rm star}$, has strong constraints on the shape of the underlying DM halo. 
}
\label{fig:modelSB}
\end{figure*}

\begin{figure*}
\centering\includegraphics[width=18cm]{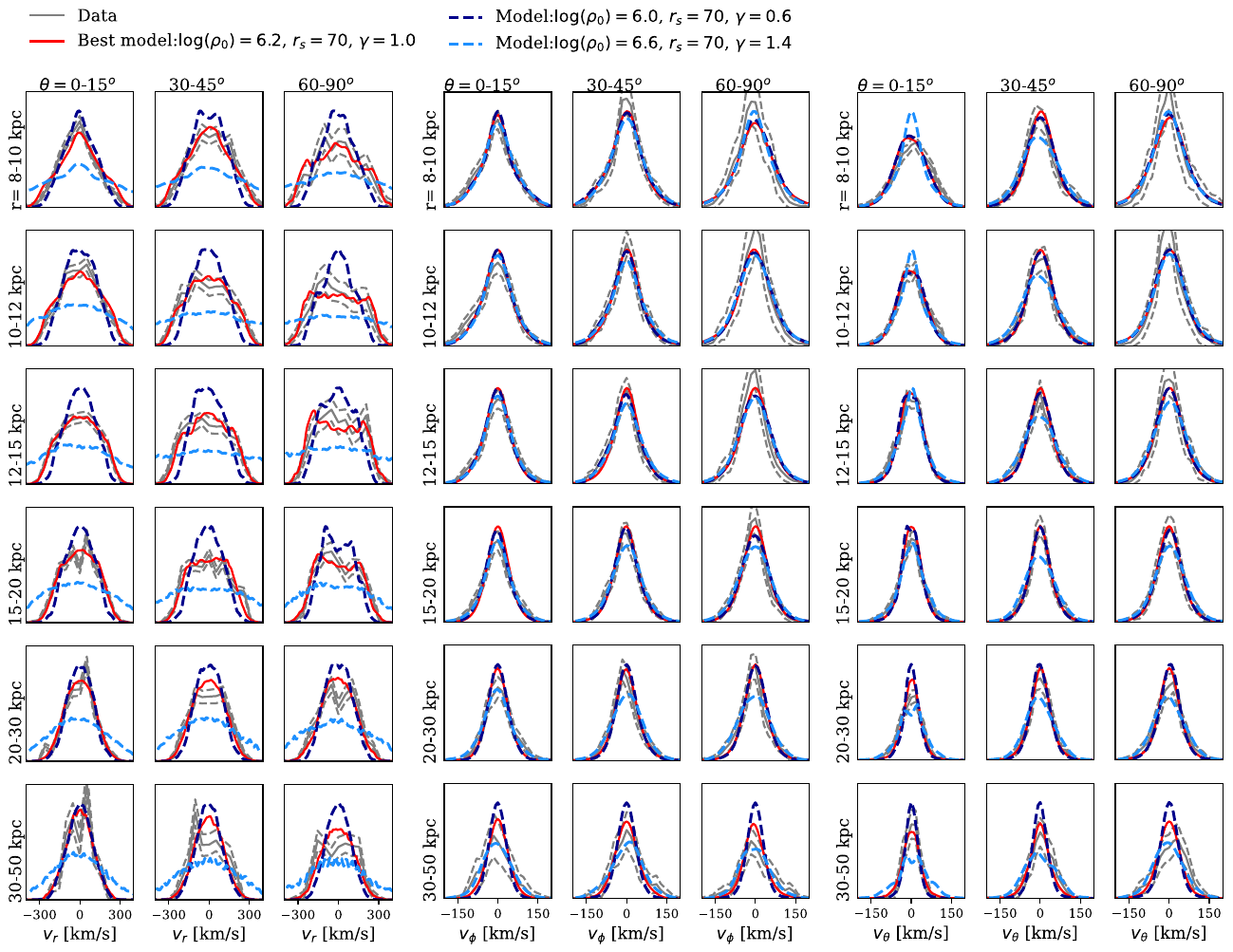}
\caption{\textbf{Comparison of the observational data and the empirical orbit superposed model: the 3D velocity distributions $v_r$, $v_\phi$, and $v_{\theta}$ in different spatial bins.} We divided the system into $7\times 5$ bins in $r$ versus $\theta$, but eliminating the first row with $r<8$ kpc and only showing the odd number of columns here. In each sub-panel, the solid and dashed grey curves are the mean and $1\sigma$ uncertainty of the distribution constructed from observational data. The red, dark blue, and light blue curves were constructed by models with different radial distributions of the underlying DM mass. The model with a DM mass of $\log(\rho_0 [M_{\odot}/{\rm kpc}^3]) = 6.2$, $r_s = 70$ kpc, and $\gamma = 1.0$ (red) matches the data well. The velocity distribution of stars, especially $v_r$, strongly constrains the radial distribution of the underlying DM distributions. 
}
\label{fig:modelv}
\end{figure*}

We investigated the parameter space of $\rho_0$, $r_s$, $\gamma$, $p_{\rm DM}$, and $q_{\rm DM}$ within the gravitational potential. A parameter grid was established with intervals of 0.1, 10 kpc, 0.02, 0.04, and 0.02 for the five parameters, respectively. An iterative approach was used to find the optimal models. The process began with an initial model, and after developing initial models, iterative refinement followed. During each iteration, we identified the optimal models using the criterion $\chi^2 - \min (\chi^2) < \chi^2_s$, where $\chi^2_s = 100$. We then generated new models by walking two steps in every direction of the parameter grid of each optimal model. This approach guided the search towards the lower $\chi^2$ within the parameter grid and halted when the model with the minimum $\chi^2$ was identified. The relatively high value of $\chi^2_s$ was selected to ensure that all models within the confidence level of $1\sigma$ were considered before the end of the iteration. Ultimately, we determined the best-fitting models by achieving the minimum $\chi^2_{\rm tot}$. 

The $1\sigma$ confidence level was determined by bootstrapping. We took the gravitational potential of the best-fitting model and perturbed the position and velocity of observed stars with their uncertainty for 100 trials. In each trial, we recomputed the model and evaluated the resulting $\chi^2_{\rm den}$, $\chi^2_{\rm vdis}$, and $\chi^2_{\rm tot}$. The standard deviation of these 100 values of $\chi^2$ was taken as the $1\sigma$ confidence level of the model $\Delta \chi^2_{1\sigma} $.

The parameter grid we explored for the Milky Way is illustrated in Fig.\ref{fig:grid} with about 15,000 models calculated. Both the density distribution, $\chi^2_{\rm den}$, and the velocity distributions, $\chi^2_{\rm kin}$, yield consistent results of the gravitational potential parameters, with $\chi^2_{\rm den}$ playing a key role in determining the DM shape parameters, $p_{\rm DM}$ and $q_{\rm DM}$. 
We established the DM halo shape parameters as $q_{\rm DM} = 0.92\pm0.08$ and $p_{\rm DM} = 0.8\pm0.2$, a nearly oblate DM halo, but with the long-intermediate axis plane vertically aligned with the stellar disc ($q_{\rm DM} > p_{\rm DM}$) that is strongly preferred by our model. 

The three parameters determining the radial distribution, $\rho_0$, $r_s$, and $\gamma$, are still degenerate. However, the enclosed DM mass profile within the data coverage is reasonably constrained. We determined the DM mass to be $M_{\rm DM}(<50 {\rm kpc}) = (5.4\pm 1.6) \times 10^{11}\,M_{\odot}$. The circular velocity at the solar location ($r=8.2$ kpc \citep{Bland-Hawthorn2016}) was found to be $V_c = 220_{-23}^{+29}$ km/s, which agrees well with previous findings \citep{McMillan2017MNRAS.465...76M}. 
Our derived enclosed mass profile and rotation curve for the Milky Way are presented in Fig.\ref{fig:mass}.

\subsection{The best-fitting models}

To illustrate how well the models match the data, we show the stellar density distribution in the $R_{\rm gc}-z_{\rm gc}$ plane in Fig.\ref{fig:modelSB}.
The stellar density distributions of the data and model were constructed in the same manner. We further extracted the flattening of the halo stars, $q_{\rm star
}$, along the iso-density contours of $\log(\rho_{\rm star}) = [-3,-2,-1,0,1,2,3]$ and compared that from the data and the model. As is shown in Fig.\ref{fig:modelSB}, the model with an underlying DM halo of $q_{
DM} = 0.92$, $p_{\rm DM} = 0.80$ constructed a stellar halo density that matches the data well. Variations in the DM halo shape yield significant deviations in the stellar density distribution of the orbit superposition model from the actual data, especially the flattening $q_{\rm star}$ at different radii. In other words, the density distribution of the halo stars, especially the flattening $q_{\rm star}$, places strong constraints on the shape of the underlying DM halo; our results for the 3D shape of DM halo are highly data-driven.

To analyse velocity distributions, we categorised model particles into bins of $N_{\rm bin} = 7 \times 5$ in $r$ and $\theta$. The divisions are $r = [4,8,10,12,15,20,30,50]$ kpc and $\theta=[0,15,30,45,60,90\degree]$. We then constructed the velocity distributions ($v_r$, $v_{\phi}$, $v_{\theta}$) within each bin for the data and model, as is illustrated in Fig.\ref{fig:modelv}.  
Likewise, the model with DM mass of $\log(\rho_0 [M_{\odot}/{\rm kpc}^3]) = 6.2$, $r_s = 70$ kpc, $\gamma = 1.0$ (red) matches the data well, while an altered density profile results in notable differences in the model's velocity distributions. The velocity distribution of stars, especially $v_r$, strongly constrains the radial distribution of the underlying DM distributions. We also note that our best-fitting model, which constructed both stellar density and the $3D$ velocity distributions, agrees with the data generally well, supporting our hypothesis of a `stationary system'. 

\subsection{Models with variable $p_{\rm DM}(r)$, $q_{\rm DM}(r)$}
The best-fitting models with constant $p_{\rm DM}$, $q_{\rm DM}$ generally succeed in replicating the full DF of the data. There is a small deviation between the 2D stellar density distribution constructed from the data and the model. This deviation is not statistically significant, but it becomes noticeable when we evaluate the flattening of the stellar halo ($q_{\rm star}$) along the iso-density contours, as is shown in the right panel of Fig.~\ref{fig:modelSB}.

We further created a set of models for the Milky Way with $p_{\rm DM}(r)$, $q_{\rm DM}(r)$ following equations~\ref{eqn:pr} to \ref{eqn:qr}. We fixed $r_q=20$ kpc and allowed $p_{\rm in}$, $p_{\rm out}$, $q_{\rm in}$, and $q_{\rm out}$ to be free parameters. The parameters determining the radial DM profile, $\rho_0$, $r_s$, and $\gamma$, do not degenerate significantly with the 3D shape. To optimise the efficiency of the model, we fixed them at $\log(\rho_0 [M_{\odot}/{\rm kpc}^3]) = 6.3$, $r_s$=40 kpc, and $\gamma=1.4$. We created about 2000 models to explore the parameter space of $p_{\rm in}$, $p_{\rm out}$, $q_{\rm in}$, and $q_{\rm out}$. We obtained models in which $q_{\rm in} <q_{\rm out}$ and $p_{\rm in} > p_{\rm out}$ are preferred, although statistically not significant. The flattening of the stellar halo ($q_{\rm star}$) of a best-fitting model is shown in Fig.\ref{fig:modelSB}, which indeed matches the data slightly better.
We extracted $p_{\rm DM}$ and $q_{\rm DM}$ at $r=15, 30, 50$ kpc from each model and compare these with cosmological simulations in the next section.

\section{Comparison with cosmological simulations}
\subsection{Shape of the dark matter halo}

\begin{figure}
\centering\includegraphics[width=8cm]{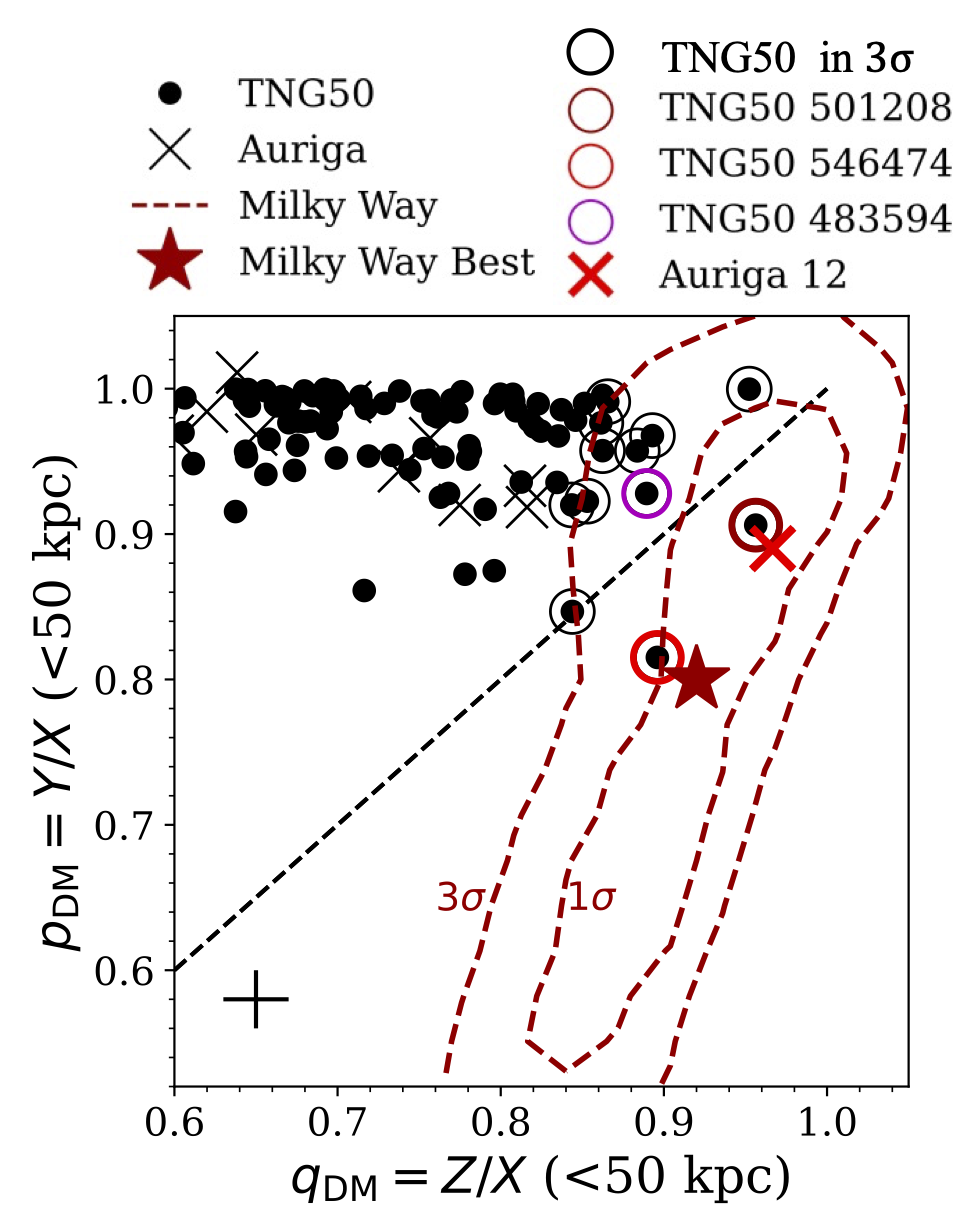}
\caption{\textbf{DM halo shape, $q_{\rm DM} = Z/X$ versus $p_{\rm DM} = Y/X$, of the Milky Way in comparison with TNG50 and Auriga galaxies.} The dashed red contours indicate the $1\sigma$ and $3\sigma$ confidence level of the Milky Way DM axis ratios obtained by our models. The red star marks the best-fitting model with $q_{\rm DM}=0.92$ and $p_{\rm DM}=0.8$. Each black dot is a Milky Way-like galaxy from TNG50 simulations. Each black multiple is a galaxy from Auriga simulations. The plus at the bottom left indicates the typical uncertainty of DM halo shapes measured for TNG50/Auriga galaxies. The four simulations highlighted by red/magenta circles/multiple are those associated with a vertical satellite plane. As we show later, those highlighted by black circles are all simulations lying within a $3\sigma$ range of the Milky Way but without significant vertical satellite planes.
}
\label{fig:pq}
\end{figure}

\begin{figure*}
\centering\includegraphics[width=18cm]{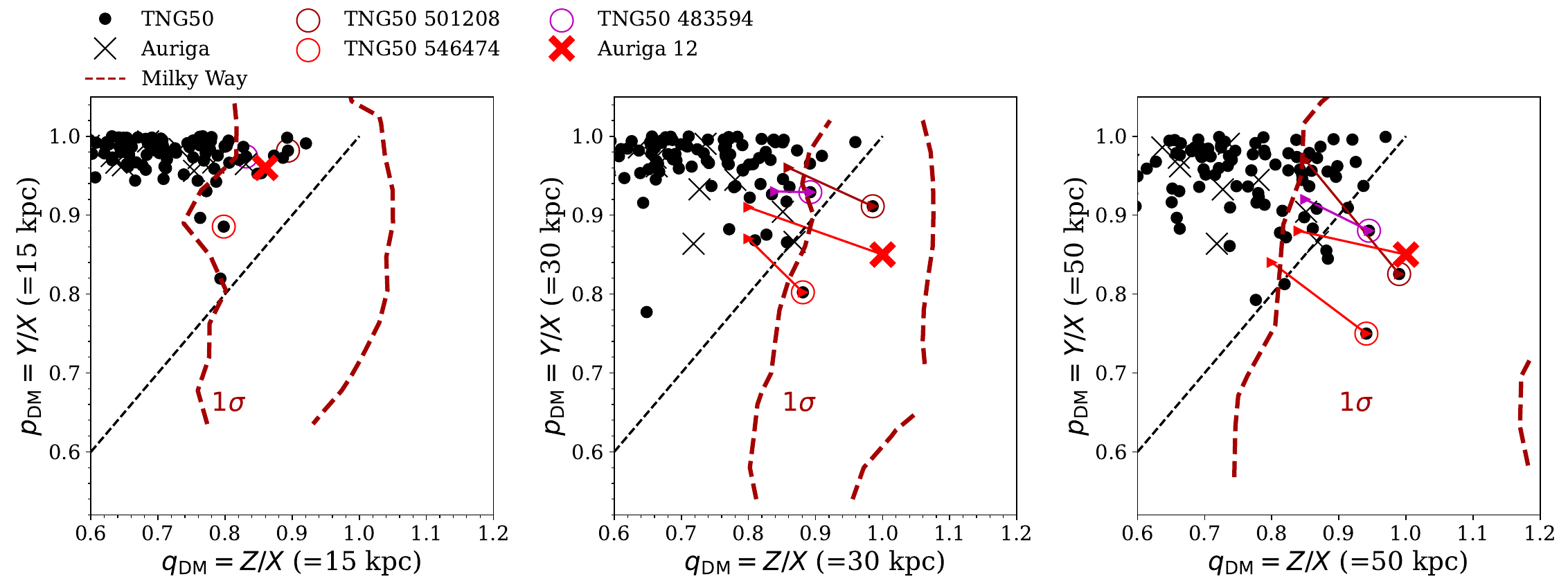}
\caption{DM axis ratio measured at three different radii, $r=15,30,50$ kpc, for TNG50 and Auriga galaxies, in comparison with that of the Milky Way obtained by our model allowing $p_{\rm DM}(r)$ and $q_{\rm DM}(r)$ vary as a function of radius. This model has a large uncertainty and we only weakly constrain the DM axis ratios at different radii, which shows a trend to be more oblate in the inner regions and to be more vertically aligned in the outer regions, consistent with the trend shown in the Milky Way analogues, TNG50 501208, TNG50 546474, Auriga 12, and TNG50 483294. For these four galaxies, the arrows point to their intrinsic axis ratios measured at the principle axes of the DM halo. These four galaxies have twisted DM halos and are vertically aligned at $r\gtrsim 20$ kpc, as we obtained for the Milky Way, bur their intrinsic DM shapes are common compared to the general TNG50/Auriga galaxies if we hypothetically tilt their disc and inner DM halo to the long-intermediate axes plane of the outer DM halo.
}
\label{fig:pq_var}
\end{figure*}

We took 198 Milky Way-like or M31-like galaxies \citep{Pillepich2024MNRAS.535.1721P} from the TNG50 simulations \citep{Nelson2019ComAC...6....2N, Nelson2019MNRAS.490.3234N, Pillepich2019MNRAS.490.3196P} and 30 galaxies from Auriga \citep{Grand2017MNRAS.467..179G}, and we performed further selections to enhance their comparability with the Milky Way. 
We calculated the flattening of the stellar component, $q_{\rm star} = z_{\rm gc}/x_{\rm gc}$, of each simulation and defined the disc fraction by considering stars with circularity $\lambda_z>0.5$ as part of the disc. Note here that we did not define a dynamically cold disc with $\lambda_z>0.7$ \citep{Genel2015ApJ...804L..40G} or $\lambda_z>0.8$ \citep{Zhu2018NatAs...2..233Z}. Instead, we included some stars on dynamically warm orbits to make the disc comparable to the Milky Way disc that was defined by fitting exponential profiles to the stellar number density distribution \citep{Grand2017MNRAS.467..179G}. The Milky Way has $f_{\rm disc} \sim 0.7$ and global $q_{\rm star} \sim 0.15$ \citep{Bland-Hawthorn2016}. We selected galaxies using a slightly loose threshold with $f_{\rm disc} > 0.6$ and $q_{\rm star} < 0.3$ to retain a relatively large sample. This selection removed galaxies that have experienced relatively recent major mergers, which create a more prominent bulge or stellar halo than that of the Milky Way. We further removed a few galaxies with ongoing major mergers that have not stabilised their stellar components. This results in 94 TNG50 galaxies and 11 Auriga galaxies. 

We calculated the intrinsic shape of the DM halo in the co-ordinate $\boldsymbol X\equiv \{X,Y,Z\}$, defined exactly the same way as we defined it for the Milky Way in our model. We first defined the $Z$ axis as being vertical to the stellar disc. Then, we projected the DM halo along the $Z$ axis and defined the long and short axes in the disc plane as $X$ and $Y$, respectively. We measured the intrinsic shape of the DM halo in the fixed co-ordinate system and measured the axis ratios $p_{\rm DM} = Y/X$ and $q_{\rm DM} = Z/X$ at $r=15,30,50$ kpc, respectively. In this definition, $p_{\rm DM}$ is restricted to be less than unity, while $q_{\rm DM}$ is allowed to be either lower or higher. 

We took the average of $p_{\rm DM}$ and $q_{\rm DM}$ measured at $r=15,30,50$ kpc for all 94 TNG50 and 11 Auriga galaxies and compared them with the Milky Way DM axis ratio obtained by our default model with constant $p_{\rm DM}$ and $q_{\rm DM}$. As is shown in Figure~\ref{fig:pq}, this particular DM halo shape of the Milky Way is not typically anticipated by the $\Lambda$CDM model for galaxy formation and is a rarity in both TNG50 and Auriga.
Nevertheless, a total of 13 out of the 105 TNG50/Auriga galaxies are found within the $3\sigma$ confidence interval of our model results for the Milky Way. The three within $1\sigma$, TNG50 501208, TNG50 546474, and Auriga 12, have intrinsically oblate DM halos, but with the long-intermediate axis plane vertically aligned with the stellar disc. 

We further show the DM axis ratios of TNG50/Auriga galaxies measured at different radii in Fig.~\ref{fig:pq_var} and compare them with the Milky Way DM axis ratio obtained by the model with varying $p_{\rm DM}(r)$ and $q_{\rm DM}(r)$. 
The above three galaxies, with average axis ratios within $1\sigma$, actually have twisted DM halos that are more oblate in the inner regions and become more vertically aligned in the outer regions. TNG50 483594, outside of $1\sigma$ but within $3\sigma$, exhibits a similar trend but is intrinsically rounder. Although not statistically significant, the Milky Way DM halo is likely to show a similar trend, as is suggested by our model with varying $p_{\rm DM}(r)$ and $q_{\rm DM}(r)$. The radial variation in the DM halo shape that we found, although tentative, is consistent with recent results from the modelling of the Sagittarius stream \citep{Vasiliev2021MNRAS.501.2279V}.

\subsection{Arrangement of the satellite system}

\begin{figure*}
\centering\includegraphics[width=18cm]{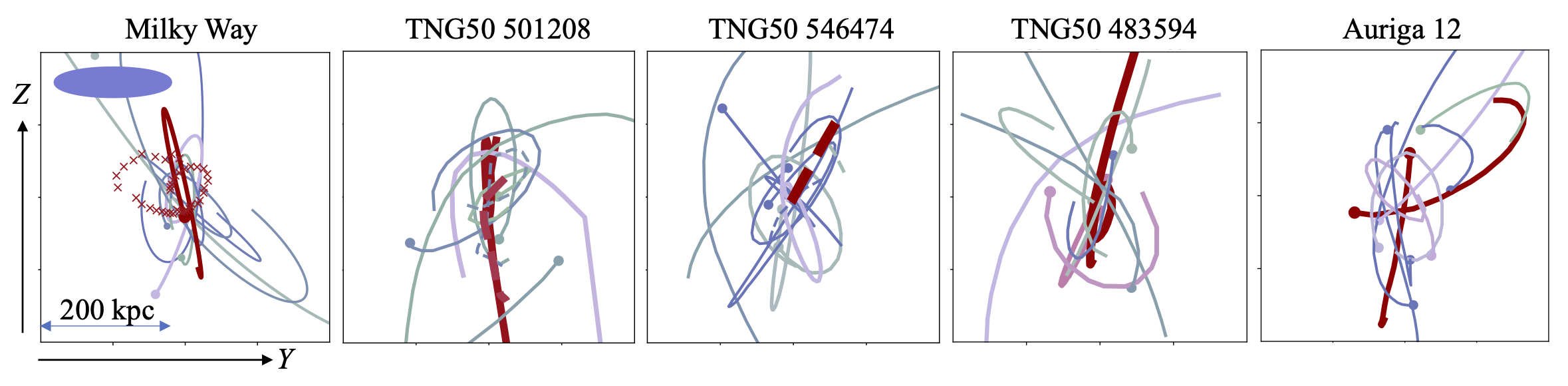}
\caption{Orbit configuration of satellite galaxies for the Milky Way and the four analogues highlighted in Fig.\ref{fig:pq} and Fig.\ref{fig:pq_var}. We show orbits of the ten brightest or most massive satellites in $YZ$ planes. The blue disc at the top left indicates the orientation of the stellar disc. The analogues exhibit striking similarities with the Milky Way that the DM halo coincides with a plane of satellites and vertically aligns with the Galactic disc. 
}
\label{fig:sat_5}
\end{figure*}

The so-called `satellite plane' of the Milky Way is especially notable in the orbits of the most massive satellites \citep{Sawala2023NatAs...7..481S}. We first illustrate the orbital configuration of the massive satellite galaxies in the Milky Way, as well as those in the 13 TNG50/Auriga galaxies whose DM shapes fall within the $3\sigma$ confidence level of the Milky Way. 

For the Milky Way, we selected the ten brightest satellites and used their observed 3D positions and velocities as initial conditions, as per \citet{Pace2022ApJ...940..136P}, to compute their orbital paths within our derived optimal gravitational potential. For each of the TNG50/Auriga galaxies, we took the ten most massive satellites and used their actual trajectories back approximately one orbital period from $z=0$ in most cases. For the satellites that have been orbited for many periods and are already very close to the galaxy centre at $z=0$, we took an orbital period from 3 Gyr ago for better visualisation.

 In Fig.\ref{fig:sat_5}, we show these four galaxies that have a nearly oblate outer DM halo vertical to the stellar disc: TNG50 501208, TNG50 546474, Auriga 12, and TNG50 483594. They exhibit vertical satellite planes similar to the Milky Way; the first three are the only galaxies with DM shapes that fall within the $1\sigma$ confidence level of the Milky Way. There are no clear satellite planes in the remaining nine galaxies within $3\sigma$ (see Appendix~\ref{ap:sat}).

In general, a correlation is evident between the DM halo shape and the arrangement of the satellite systems in TNG50 galaxies (see Appendix~\ref{ap:sat}). Most TNG50 galaxies have oblate DM halos with long-intermediate axes aligned with the stellar disc and paired with coplanar satellite systems. While galaxies with almost spherical DM halos display satellites distributed in a nearly isotropic fashion, those with DM halos aligned vertically to the stellar disc tend to have satellite systems organised vertically as well.
This correlation is expected as a result of the DM halo assembling over time through the accretion of satellite galaxies with similar net angular momentum, consistent with previous findings from other cosmological simulations \citep{Shao2016MNRAS.460.3772S}.

\subsection{A vertically orientated dark matter halo in alignment with a satellite plane results from a flip of the stellar disc}
\begin{figure*}
\centering\includegraphics[width=18cm]{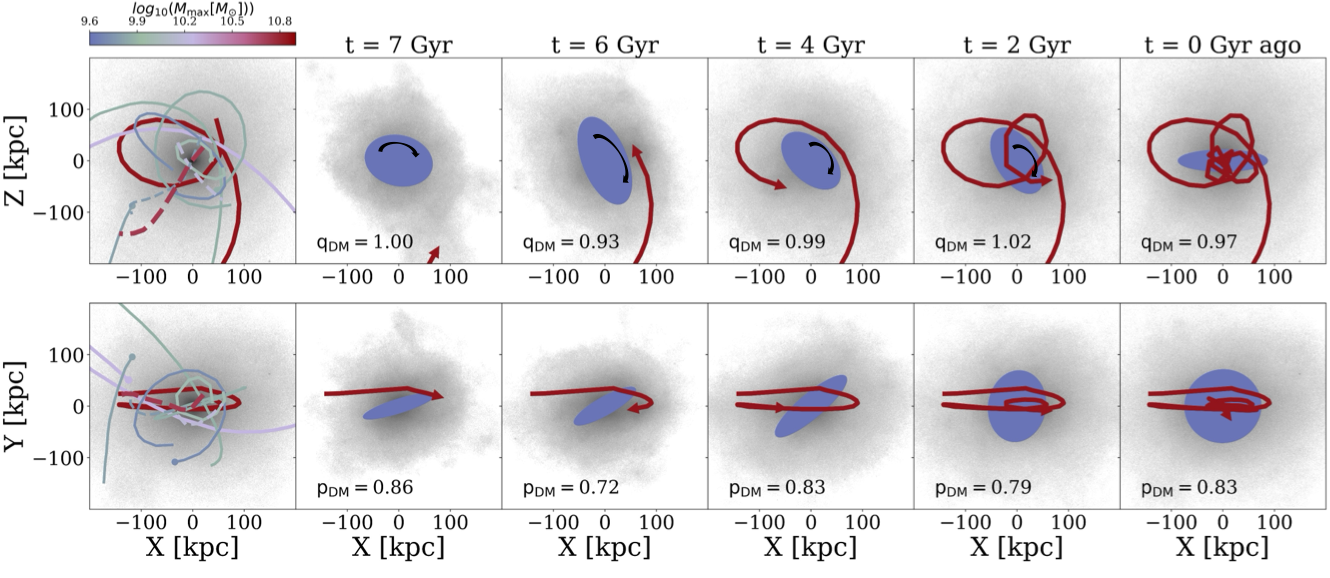}
\caption{\textbf{Spiral-in minor merger flips the stellar disc illustrated by TNG50 501208.} The left column show orbits of the ten most massive mergers coloured their maximum history mass. The dashed red curve with $M_{\rm max}=5.3\times10^{10}$\Msun\, is a major merger finished at about 8.5 Gyr ago, and the solid red curve with $M_{\rm max}=7.1\times10^{10}$\Msun\, is the most massive minor merger followed. The rest columns show the DM distribution (the grey shadow), the projection of the stellar disc (the blue ellipse with size enlarged five times for visualisation), and the orbit progress of the most massive minor merger (solid red curve) at different snapshots in the history. The axis ratios of the DM halo measured at $r=50$ kpc are labeled. A thin stellar disc was formed right after the major merger finished at about 8.5 Gyr ago. At about $t=6-7$ Gyr ago the disc was aligned almost co-planar with the long-intermediate axis plane of the DM halo and orbital plane of the satellites. The disc was flipped in the last seven gigayears, associated with the spiral-in minor merger (solid red curve), while the general shape of DM halo (illustrated by $q_{\rm DM}$ and $p_{\rm DM}$ measured at 50 kpc in a fixed co-ordinate) and satellite arrangement remain unchanged, resulting in the stellar disc vertical to the DM halo and satellite orbit plane at present (also see Movie S1). }
\label{fig:501208}
\end{figure*}

\begin{figure*}
\centering\includegraphics[width=8cm]{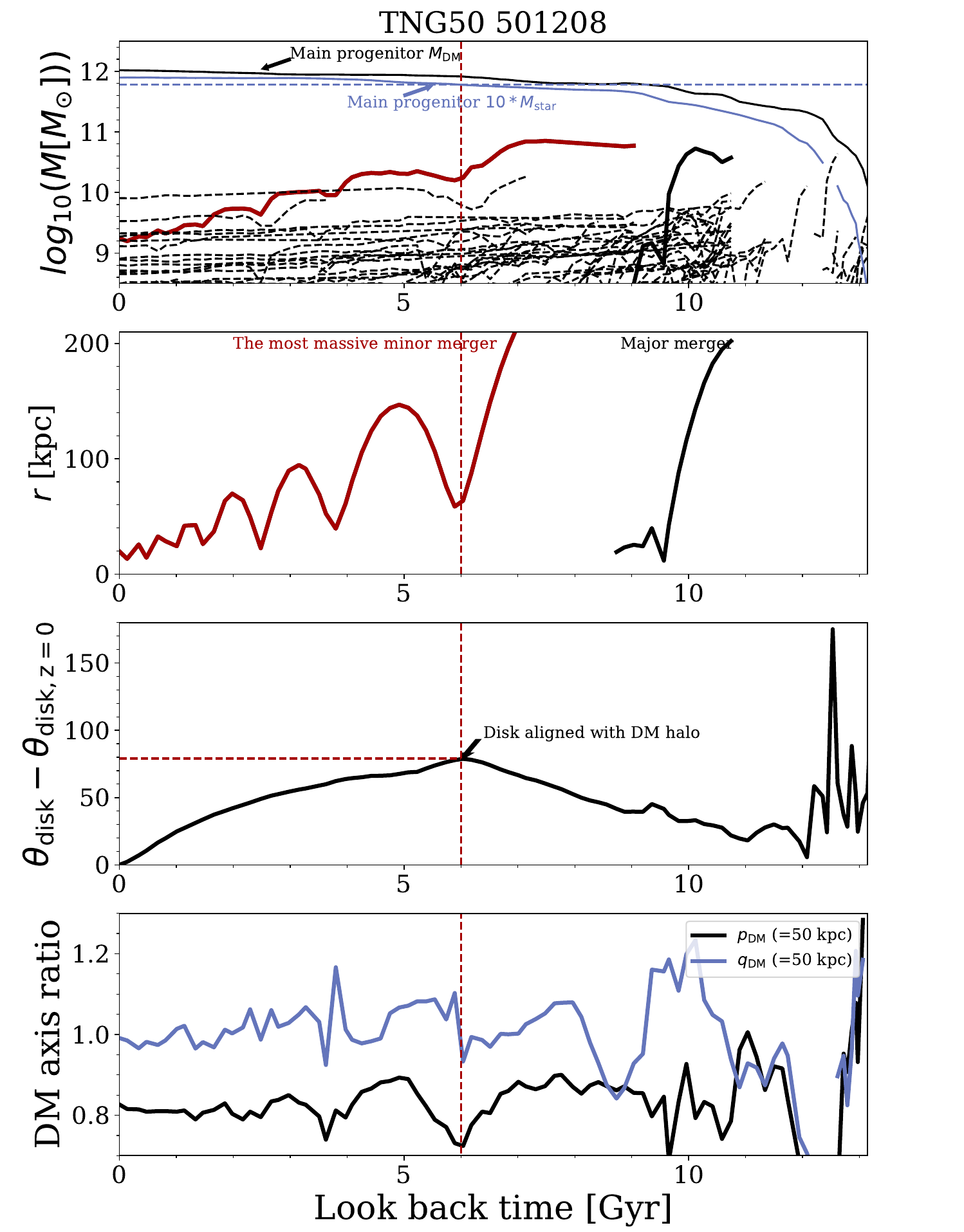}\centering\includegraphics[width=8cm]{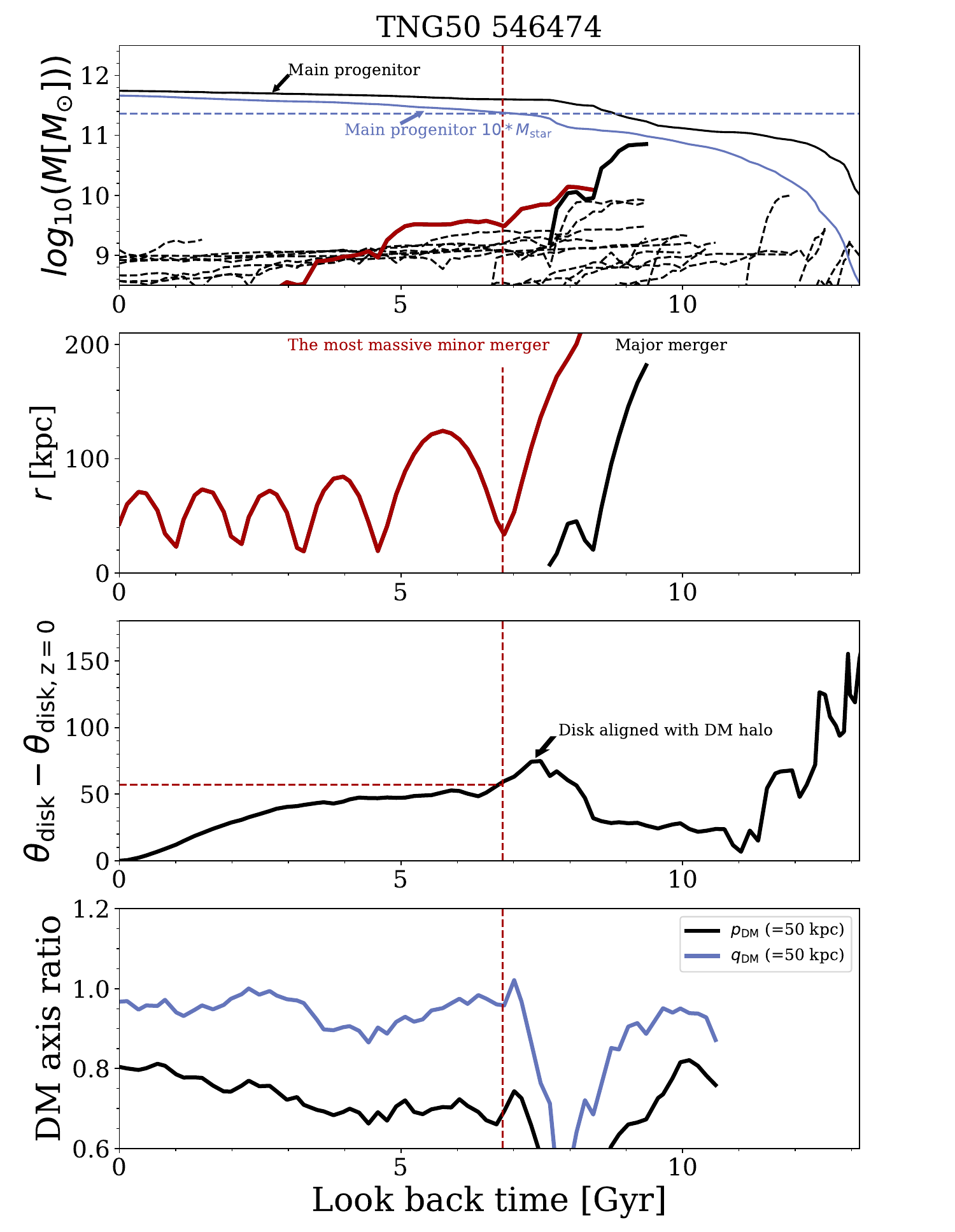}
\caption{Evolution history of TNG50 501208 (left) and TNG50 546474 (right). We describe the four panels with TNG50 501208. {\bf Top:} Assembly history of the galaxy. The solid thick curves represent two of the most massive mergers the galaxy experienced. The dashed curves represent small mergers. TNG50 501208 has assembled $80\%$ of its DM mass and $72\%$ of its stellar mass at $t=6$ Gyr ago. {\bf Second row}: Distance of the satellite to the galacetic center as a function of time. The solid black curve is a major merger ended at $t \sim 8.5$ Gyr ago. The red curve is the following most massive minor merger with $M_{\rm max} = 7.1\times 10^{10}$\,\Msun. The vertical dashed red line indicates the time at which it reached the pericenter for the first time. {\bf Third row:} Variation in the disc spin direction as a function of time. The disc was co-aligned with the long-intermediate axis plane of the DM halo at $t\sim 6 $ Gyr ago, and it has tilted $80\degree$ in the past 6 Gyr. {\bf Bottom:} Evolution of the DM axis ratios, with $p_{\rm DM}= Y/X$ and $q_{\rm DM}=Z/X$. The DM shape remain largely unchanged in the past 8 Gyr. The four panels in the right show a similar evolution history of TNG50 546474.} 
\label{fig:501208r}
\end{figure*}

\begin{figure*}
\centering\includegraphics[width=8cm]{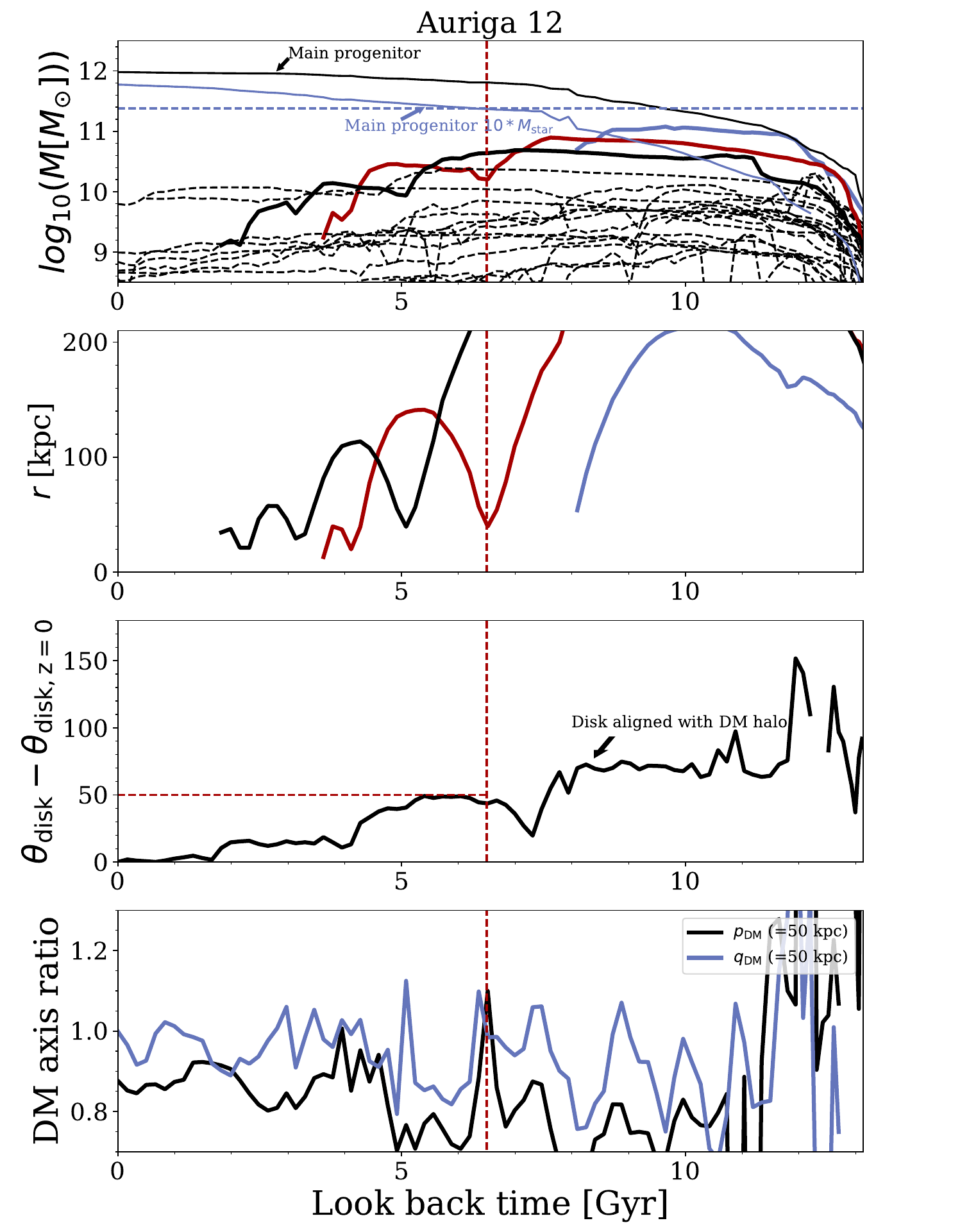}\centering\includegraphics[width=8cm]{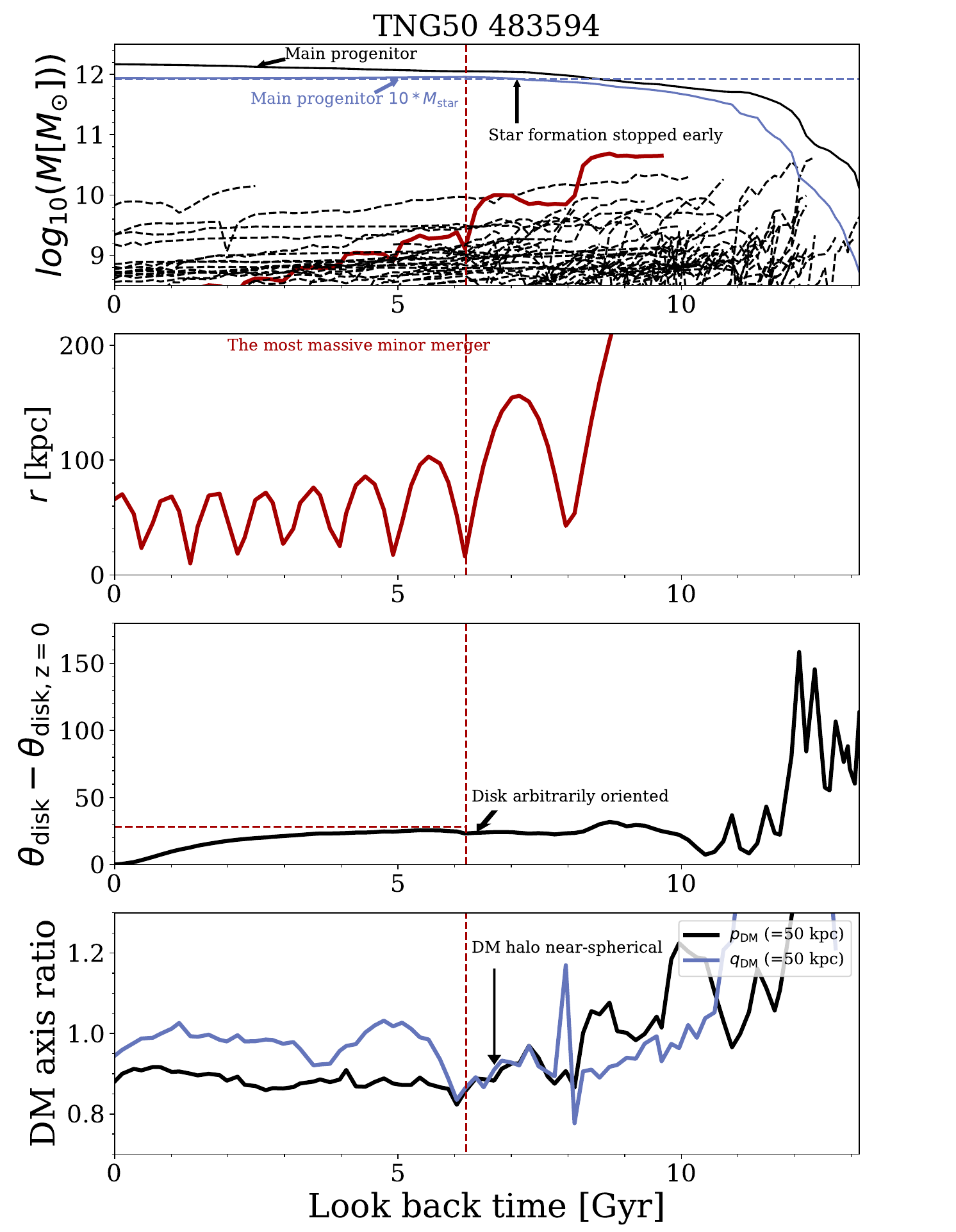}
\caption{Similar to Fig.\ref{fig:501208r}, but for Auriga 12 (left) and TNG50 483594 (right). 
}
\label{fig:Au12r}
\end{figure*}

We further traced the evolutionary history of the four Auriga/TNG50 galaxies, TNG50 501208, TNG50 546474, TNG50 483594, and Auriga 12, and found that the distinct vertical alignment of the DM halo and satellite plane is due to the tilt of the stellar disc with a large angle, which we call a `flip of the disc'. To check the evolutionary history, we paid special attention to the merger history, the evolution of the stellar disc, and the shape of the DM halo.
We calculated the spin direction of all stars at $r<20$ kpc at each snapshot in the history and defined $\theta_{\rm disc}$ as the angle between the spin vector and its projection line in the XY plane. In this definition, we have $\theta_{\rm disc, z=0} = 90\degree$. We then calculated the DM axis ratios at 50 kpc ($p_{\rm DM} = Y/X$, $q_{\rm DM} = Z/X$) in different snapshots. $XYZ$ is the co-ordinate defined at $z=0$, with the $Z$ axis perpendicular to the disc, as we defined previously.

As is illustrated in Fig.\ref{fig:501208} and quantified in Fig.~\ref{fig:501208r}, the last major merger of TNG50 501208 ended $t\sim 8.5$ Gyr ago. A stellar disc was formed gradually after the major merger. At $t\sim 6-7$ Gyr ago, the stellar disc was almost aligned in the long-intermediate axis plane of the DM halo; if not perturbed by other mergers, the system would stabilise in this configuration.  
The galaxy was followed by a series of minor mergers, the most massive one having $M_{\rm max} = 7.1\times 10^{10}$\,\Msun coming in on an orbit that is retrograde to the disc rotation. At $t = 6$ Gyr ago, when the massive minor merger reached its pericenter for the first time, the disc started to tilt, initiating continuous tilting in the subsequent 6 Gyrs.
The disc has since flipped about $80\degree$ until $z=0$, with an average rate of $\sim 13\degree$ per Gyr, while it was not significantly perturbed. Although the satellite perturbed the DM halo's shape at pericenter, the DM halo quickly re-stabilised. The mass accreted from this satellite and other minor mergers continuously built the DM halo along the orbit, consistent with the pre-existing DM halo shape; the axis ratios of the DM halo measured at $r=50$ kpc remained largely unchanged. 
The disc flip resulted in the DM halo and stellar disc's vertical alignment at $z=0$. 

The satellite galaxy that spiralled inwards brings a substantial net angular momentum. It is important to note that a tilt rate of $\sim 13\degree$ per Gyr corresponds to about 2 km/s at a radius of $r=10$ kpc, indicating a very minor rigid motion of the disc. This scenario requires torques that transfer less than $\lesssim 1\%$ of the angular momentum of the satellite galaxy, which has a total mass comparable to that of the stellar disc. In our analysis of 13 TNG/Auriga galaxies within a $3\sigma$ threshold, eight galaxies exhibit significant disc tilts, with tilt angles ranging from $30\degree$ to $100\degree$. Seven of them are likely caused by minor mergers with $M_{\rm max} > 10^{10}\,M_{\odot}$, mostly spiralled in (see appendix~\ref{ap:sat}). Disc flips are fairly common in cosmological simulations \citep{Dillamore2022MNRAS.513.1867D,Earp2017MNRAS.469.4095E,Bett2012MNRAS.420.3324B}. Direct N-body simulations show that the torque on the stellar disc from a satellite galaxy is a promising mechanism that induces the disc tilt \citep{Dodge2023MNRAS.518.2870D}. Previous analysis of Auriga galaxies shows that disc tilt is correlated with the time duration of close satellite interactions \citep{Gomez2017MNRAS.472.3722G}, consistent with our results. It is also found that even low-mass satellites that penetrate the outer regions of a galaxy can significantly perturb and tilt a host galactic disc \citep{Gomez2017MNRAS.465.3446G}, due not only to direct tidal perturbation but also to the generation of asymmetric features in the DM halo that can be efficiently transmitted to its inner regions.

In a similar pattern, the evolution of TNG50 546474 is presented in the right panel of Fig.~\ref{fig:501208r}. This galaxy underwent a major merger that ended around $t\sim 7.5$ Gyr ago, immediately followed by the most massive minor merger with $M_{\rm max}=1.4\times10^{10}$\Msun. At $t\sim 7$ Gyr ago, after the reformation of the disc post-major merger and stabilisation of the DM halo, the disc and DM halo were nearly coplanar and aligned. Subsequently, the DM halo's shape saw minimal change, although it gained about $30\%$ of its mass in the upcoming 7 Gyr through minor mergers on orbits consistent with the pre-existing DM halo shape. At the same time, the disc has tilted approximately $60\degree$ with an average rate of $\sim 8\degree$ per Gyr.

Similar figures for Auriga 12 and TNG50 483594 are shown in Figure~\ref{fig:Au12r}.
Auriga 12 has had more frequent recent mergers. Its stellar disc was coplanar aligned with the DM halo at $t\sim 8-10$ Gyr ago; the torques from the combination of a few subsequent minor mergers have tilted the disc by $\sim 90\degree$. The DM halo was influenced but has maintained a similar shape since about 10 Gyr ago.
 
 The DM shape of TNG50 483594 is outside the $1\sigma$ but within the $3\sigma$ confidence level of the Milky Way. This galaxy is special in that it was quickly assembled with many head-on mergers at $t\gtrsim 8$ Gyr ago, but without a GSE-like major merger, which resulted in a rather spherical DM halo. Due to frequent mergers in the early Universe, it has quickly assembled stellar mass and halted star formation. With a near-spherical DM halo, its stellar disc was arbitrarily orientated at $t=6-10$ Gyr ago. The subsequent minor mergers exhibit a satellite plane whose contribution has not significantly changed the DM halo shape until $t\sim 6$ Gyr ago, along with a massive minor merger with $M_{\rm max} = 4.5\times 10^{10}$\Msun. The minor mergers built the halo along their orbits and tilted the disc by about $30\degree$ in another direction, leading to the vertical alignment of the DM halo and the disc at $z=0$. 
  In this case, the tilt of the disc is not the only reason; it strengthens the misalignment and contributes to the vertical alignment of the stellar disc and the satellite distribution observed at $z=0$. Even without the disc tilt, there is still a misalignment of the stellar disc and the satellite plane due to variations in the filament network along with mass accreted into the galaxy, as is shown in \citep{Shao2021MNRAS.504.6033S}.

Despite the tilting of the disc, the DM halo of the inner regions ($r\lesssim 20$ kpc) was tilting with the stellar disc. As a result, the DM halos of these galaxies tend to be twisted, being oblate in the inner regions, and more vertically aligned with the disc in the outer regions, as has already been shown in Figure~\ref{fig:pq_var}, consistent with the trend we obtained for the Milky Way. 

The three $1\sigma$ galaxies, TNG50 501208, TNG50 546474, and Auriga 12, would transform into a typical galaxy with an oblate DM halo by hypothetically tilting their stellar disc (and inner DM halo at $r\lesssim 20$ kpc) back to the long-intermediate axis plane of the DM halo in the outer regions.
Note that TNG50 403894 also has a twisted DM halo; however, its transition radius, similar to other cases of twisted halos caused by variations in filament accretion \citep{Shao2021MNRAS.504.6033S}, is larger than the three cases caused purely by the tilt of the disc (see Fig.\ref{fig:pq_var}).

\section{Discussion}
\subsection{Significance of the results}
\begin{figure}
\centering\includegraphics[width=9cm]{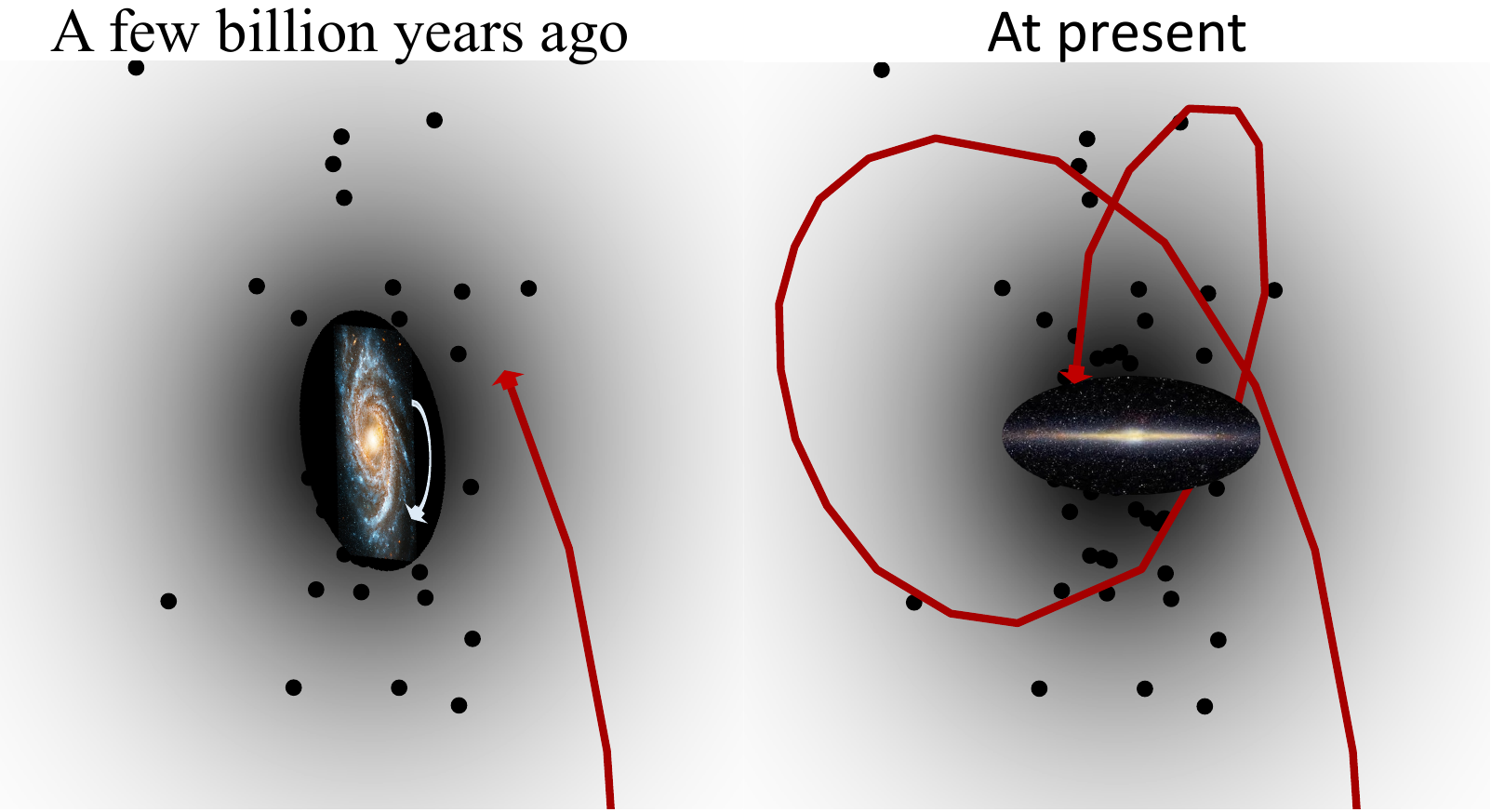}
\caption{{\bf Diagram illustrating the structural evolution that we have proposed for the Milky Way.} The grey ellipse shows the DM halo, with the darker region representing the inner halo associated with the stellar disc. The black dots denote satellite galaxies, while the red curve represents orbit of a massive satellite retrograde to the rotation of stellar disc. Several gigayears ago, the Milky Way featured an oblate DM halo, with the stellar disc, DM halo, and satellite galaxies co-aligned. Around 6-8 Gyr ago, the disc and inner halo began tilting, and this tilt continued over the subsequent gigayears, eventually leading to a twisted DM halo, with the outer DM halo and satellite plane becoming perpendicular to the stellar disc. }
\label{fig:mw_disk}
\end{figure}

In our default model with a constant axis ratio of $q_{\rm DM} = Z/X$ and $p_{\rm DM} = Y/X$, the axis ratio $q_{\rm DM}$ is constrained to be $0.92\pm 0.08$, consistent with most previous studies \cite[e.g.,][]{Wegg2019MNRAS.485.3296W, Huang2024NatAs...8.1294H, Zhang2025}. However, we have also shown that the Milky Way DM halo is triaxial, with $q_{\rm DM}>p_{\rm DM}$ strongly preferred. Although there is still some degeneracy between the two parameters, with a relatively large uncertainty on the axis ratio $p_{\rm DM}$ from our dynamical model, the allowed region in $q_{\rm DM}$ versus $p_{\rm DM}$ is already narrow enough for us to distinguish the galaxy formation scenarios.
Except for the four Milky Way analogues aforementioned, TNG50 501208, TNG50 546474, TNG50 483594, and Auriga 12, all the other TNG50 galaxies fall within the $3\sigma$ confidence level of our model, displaying nearly isotropic or coplanar satellite distributions, as was expected from their intrinsic near-spherical or normally oblate DM halos (see Appendix~\ref{ap:3sigma}). 

The presence of a vertical satellite plane thus helps to pin down the shape of the DM halo: a nearly oblate but vertically orientated DM halo is the only type allowed within the $3\sigma$ threshold of the Milky Way that is associated with a vertical satellite plane. 
In three of the four galaxies, the vertical alignment of the DM halo and satellite plane is purely due to the identical physical mechanism: a flip of the stellar disc, likely torqued by minor mergers that spiralled inwards. They are also the only three galaxies with DM halo shapes lying within $1\sigma$ range of the Milky Way.
In the remaining case, TNG50 483594, its vertical alignment was initialised by variations in the filament network along with mass accreted into the galaxy, but was strengthened by disc tilt. Its DM halo is intrinsically rounder, out of $1\sigma$ range of the Milky Way. What is more, it has a very different assembly history from the Milky Way.
We therefore strongly propose that the Milky Way arises from the same scenario as the three $1\sigma$ galaxies; otherwise, a system combining such a DM halo shape and vertical satellite plane would be highly impossible in either the TNG50 or Auriga simulations.

The merger history of the Milky Way closely resembles that of two of the aforementioned $1\sigma$ galaxies, TNG50 501208 and TNG50 546474. It is believed that the stellar disc formed after the last major merger \citep{Belokurov2020MNRAS.494.3880B} -- the Gaia-Enceladus-Sausage (GSE) merger -- occurred 8-10 Gyr ago \citep{Helmi2018Natur.563...85H, Belokurov2018MNRAS.478..611B}, comparable to the major mergers that happened in these two galaxies 9-10 Gyr ago (highlighted in Fig~\ref{fig:501208r}).
The prominent Sagittarius stream \citep{Majewski2003ApJ...599.1082M, Belokurov2006ApJ...642L.137B} imprints several spiral-in orbital passages in the halo over the past 6-8 Gyr \citep{Dierickx2017ApJ...836...92D}. Recent studies suggest that the progenitor mass of Sagittarius is likely to be $5-6\times 10^{10}$\Msun\, or even larger \citep{Gibbons2017MNRAS.464..794G, Read2019MNRAS.487.5799R}, comparable to the minor mergers that induced the disc flip in TNG50 501208 and TNG50 546474. 

Direct N-body simulations \citep{Dodge2023MNRAS.518.2870D} show that the GSE could effectively tilt the Galactic disc, if not destroy it. Although GSE-like major mergers usually result in the co-alignment of the disc and DM halo \citep{Dillamore2022MNRAS.513.1867D}, the ultimate orientation of the disc depends on the orbital properties of the satellite. We find that satellites coming in on prograde orbits usually result in the co-alignment of the disc and orbital plane, while satellites coming in on retrograde orbits could result in a disc vertical to the satellite orbital plane (Cai \& Zhu et al., in preparation). The latter is consistent with the case in the MW that GSE remnants are with retrograde rotations \citep{Helmi2018Natur.563...85H}.
We suggest that Sagittarius could also be a promising candidate that induced the disc tilting to vertical alignment, similar to the tilt that began about 6 Gyr ago in TNG50 501208. In real cases, the torques of multiple mergers could have worked together, along with the accretion of unbound DM \citep{Genel2010ApJ...719..229G}.

A diagram is included in Fig.\ref{fig:mw_disk} to illustrate the structural evolution we propose for the Milky Way. 
The current disc-halo orientation of the Milky Way may not be fully stable; the disc could still be tilting. Although there are currently no direct observational constraints, the model of the stream Pal 5's tidal tails favours a tilting rate of $15\degree$ per Gyr for the Milky Way disc \citep{Nibauer2024ApJ...969...55N}, supporting our scenario.

\subsection{The effects of a tilting disc on dynamical models}
If the Galactic disc is still tilting at $z=0$, a tilting disc might affect our dynamical models in two aspects: first, the overall potential is evolving, and thus the stellar halo, especially the inner halo, is not necessarily in a stationary state; second, the inner halo may be tilting with the disc and cause a velocity shift and finally, a twist compared to the outer halo, as suggested by the simulations. We evaluate the significance of these effects, considering a disc tilt rate of $15\degree$ per Gyr.

The cross time of stars in the inner stellar halo is 0.1-0.2 Gyr. During one period of cross time, the disc could tilt about two degrees, which has negligible effects on the potential. In other words, the potential evolves such that the motion of stars will be affected, but the evolution is so slow that the halo stars can update their motions efficiently to reach dynamical equilibrium in a timely manner. Our assumption of dynamical equilibrium should still hold with any momentum. Moreover, the dynamical equilibrium of the system is indeed supported by the good match between the data and our best-fitting model. 

If there is a velocity shift between the inner and outer halo caused by the tilt of the inner halo with the disc, it will lead to some non-equilibrium features in the outer halo. A disc tilt rate of $15\degree$ per Gyr equals 0.05 mas/yr, and corresponds to 8 km/s at $r=30$ kpc and 13 km/s at $r=50$ kpc, which are less than the net velocities we subtracted for the correction of LMC effects. The effects of tilt to the outer halo, if they exist, will be mixed with the LMC effects we subtracted. As we tested, imposing the LMC correction or not does not make a noticeable difference in our final results. We thus do not expect a noticeable difference in our final results if there is a tilt of the inner halo with the disc. In summary, the disc tilt, if it is still ongoing at $z=0$, is actually very slow and has minor effects on the kinematics of the stellar halo, which should not affect the results of the dynamical models.

\subsection{The tilt angle of the dark matter halo}
We fixed the tilt angle of the DM halo, $\beta_{q}$, to be either $0\degree$ or $90\degree$ compared to the disc, by fixing the Z axis as perpendicular to the stellar disc. Our current approach allows for a fair comparison with the cosmological simulations by defining exactly the same co-ordinates for measuring the DM shape of the simulated galaxies. In reality, the tilt angle of the DM halo is not likely to be exactly $0\degree$ or $90\degree$;
a disc vertical to the intermediate axis of the DM halo was suggested to be unstable, while a disc can persist off one of the principal planes of the potential \citep{Debattista2013MNRAS.434.2971D}.
However, the disc tilt angle is strongly degenerated with the axis ratio $q_{\rm DM}$ for the measurement.
A tilt angle of $\beta_q\sim 15 \degree$ of the DM halo compared to the stellar disc was preferred in creating the disc warp \citep{Han2023NatAs...7.1481H} and the presence of a tilted stellar halo dominated by GSE \citep{Han2022ApJ...934...14H}. 
The tilted stellar halo was used to constrain an orbit-superposition model \citep{Dillamore2025arXiv251000095D}, in which the DM tilt angle is allowed to be free. They obtain an average tilt angle of $\beta_q\sim 45\degree$ for the DM halo, but there is a strong degeneracy between the two parameters such that $\beta_q\sim 30-80\degree$ and $q_{\rm DM} \sim 0.7-0.95$ are allowed within their $1\sigma$ confidence level. Note that all the above results are mainly constrained by GSE at $r\lesssim 30$ kpc. A tilted DM halo in the inner regions is generally consistent with our results; we find tentative evidence for a radially varying shape of the DM halo, which being less tilted in the inner regions and becomes vertically aligned in the outer regions, consistent with \citet{Vasiliev2021MNRAS.501.2279V}. Data with larger spatial coverage are needed to further constrain the twist of the DM halo. Interestingly, a recent study has constructed the 3D distribution of the stellar halo out to about 50 kpc using Desi observations, which found that the Milky Way stellar halo is twisted and becomes vertical to the stellar disc in the outer regions \citep{Li2025arXiv251201350L}, which turns out to be consistent with the shape of the DM halo we found. 
\section{Summary}
We determined the 3D DM distribution of the Milky Way by utilising the empirical triaxial orbit-superposition model applied to the 6D phase-space data of halo stars provided by LAMOST and Gaia. This model accommodates a highly flexible triaxial DM halo, where the axis ratios may change with varying radii. Both the radial density profile and the 3D shape of the DM halo are constrained simultaneously. Furthermore, we benchmark our findings against 105 galaxies resembling the Milky Way from cosmological simulations. Our major findings are as follows.

\begin{itemize}
\item We discover a triaxial, nearly oblate DM halo with $q_{\rm DM} = Z/X= 0.92\pm0.08$, $p = Y/X= 0.8\pm0.2$ averaging $r\lesssim 50$ kpc, where the $Z$ axis is defined perpendicular to the stellar disc. The axis ratio $q_{\rm DM} > p_{\rm DM}$ is strongly preferred; the long-intermediate axis plane of the DM halo is unexpectedly vertical to the Galactic disc. 
\item We determine the DM mass to be $M_{\rm DM}(<50 {\rm kpc}) = (5.4\pm 1.6) \times 10^{11}\,M_{\odot}$, and the circular velocity at the solar location to be $V_c = 220_{-23}^{+29}$ km/s, which agrees well with previous research. 
\item This particular DM halo shape of the Milky Way is not typically anticipated by the $\Lambda$CDM model for galaxy
formation and is a rarity in both TNG50 and Auriga. Among the 105 galaxies in TNG50/Auriga, only 13 are within the $3\sigma$ confidence level of our Milky Way findings, and just three fall within the $1\sigma$ range. Notably, all three $1\sigma$ galaxies display a plane of satellites that aligns with the long-intermediate plane of the DM halo and is perpendicular to the stellar disc, mirroring the Milky Way.
\item This striking configuration of the DM halo and satellite system strongly suggests that the Galactic disc has flipped, torqued by minor mergers, from its original alignment with the DM halo and satellite plane, as is evidenced by Milky Way analogues we found in Auriga and TNG50.
\item We find tentative evidence that the Milky Way DM halo is twisted, consistent with alignment with the disc in the inner $r\lesssim 20$ kpc, and becomes vertically orientated in the outer regions, consistent with the prediction of disc flip scenario.
\end{itemize}  

With the 3D shape of the DM halo uncovered, we establish a coherent scenario within the $\Lambda$CDM framework for the origin of the vertically orientated DM halo and the satellite plane of the Milky Way as a result of the disc flip. This addresses the vertical alignment of the satellite plane problem in a nontrivial but straightforward way. At this point, our Milky Way is special but not an outlier in the $\Lambda$CDM framework.
Additionally, the DM shape and disc orientation have broad implications; they potentially impact the development of other Milky Way formations, including disc warps \citep[e.g.,][]{Yaaqib2025arXiv250104095Y} and stellar streams \citep[e.g.,][]{Nibauer2024ApJ...969...55N}. Our work may trigger a significant shift in understanding the formation and evolution of the Milky Way.

\begin{acknowledgement}
The authors thank David Hogg, Jianhui Lian, Zhaozhou Li, Wenting Wang for useful discussions. The authors acknowledge the support from the National Key R\&D Program of China No. 2025YFF0511002(LZ), 2022YFF0503403 (LZ), 2022YFA1602903 (XK), and No. 2024YFA1611902 (X.X.X), the CAS Project for Young Scientists in Basic Research under grant No. YSBR-062 (LZ, X.X.X, LZhang),  and the science research grants from the China Manned Space Project with NO. CMS-CSST-2025-A11 (X.X.X, LZ, CY, LZhang), the Strategic Priority Research Program of Chinese Academy of Sciences grant No. XDB1160102 (X.X.X), and National Natural Science Foundation of China (NSFC) No. 12588202 (X.X.X).

\end{acknowledgement}

\bibliographystyle{aa}  
\bibliography{ms_mw} 

\begin{appendix}
\section{Quantifying the spatial arrangement of satellites}
\label{ap:sat}

We further quantify the distribution of the overall population of satellite galaxies for the 94 selected Milky Way-like galaxies from TNG50 and the 11 galaxies from Auriga. 
We include all satellites at $z=0$ that are more massive than $10^8$\,\Msun\ and located within 200 kpc of the halo, and we characterise the arrangement of the satellite system at $z=0$ by defining two parameters: the fraction of satellites in vertically aligned orbits, denoted as $f_{\rm Sat, vertical}$, and the anisotropy of their distribution in the azimuthal angle, represented by $\xi_{\rm Sat, \phi}$.

\begin{figure}
\centering\includegraphics[width=8cm]{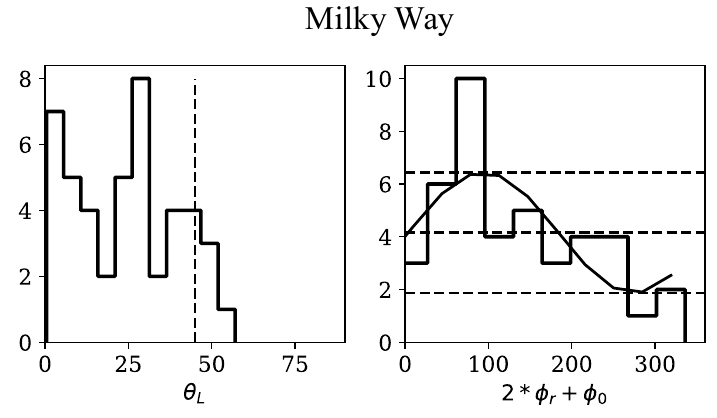}
\centering\includegraphics[width=8cm]{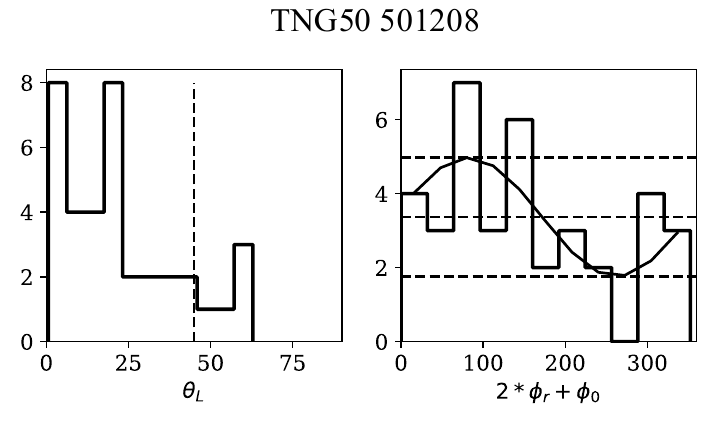}
\caption{The satellite distribution in $\theta_L$ and $\phi_r$ for the Milky Way (Top) and TNG50 501208 (Botton) illustrating the definition of $f_{\rm Sat, vertical}$ and $\xi_{\rm Sat, \phi}$. The parameter $f_{\rm Sat, vertical}$ is defined as the fraction of satellites with $\theta_L<45\degree$.  
We fit the distribution of $\phi_r$ with a sine function $n = n_1\sin(2\phi_r + \phi_0) + n_0$. The parameter $\xi_{\rm Sat, \phi} = n_1/n_0$ is defined to describe the anisotropy of satellite galaxies distributed in azimuthal angle $\phi_r$. 
}
\label{fig:qfsat}
\end{figure}

 First, we compute their angular momentum; the orientation of the angular momentum $\theta_L$ is determined by the angle between the spin direction and its projection line on the disc plane. A satellite with $\theta_L \approx 0$ is in an orbit vertically aligned with the stellar disc. The $f_{\rm Sat, vertical}$ is defined as the fraction of satellites with $\theta_L < 45\degree$. We evaluated the azimuthal anisotropy of the satellite distribution by calculating their azimuthal angle $\phi_r$ in the disc plane. The distribution of all satellites' $\phi_r$ is plotted, and we fit the histogram with a sine function $n = n_1\sin(2\phi_r + \phi_0) + n_0$. An anisotropy parameter, $\xi_{\rm Sat, \phi} = n_1/n_0$, is defined. The definition of these two parameters is illustrated in Fig.~\ref{fig:qfsat}. Under this definition, a perfectly isotropic satellite system has $f_{\rm Sat, vertical} = 0.707$ and $\xi_{\rm Sat, \phi} = 0$. 
 
  For the Milky Way, we took 52 satellite galaxies with measured 3D position and velocity \citep{Pace2022ApJ...940..136P}, and added LMC to the data, thus with 53 in total. As shown in Figure~\ref{fig:qfsat}, the satellite galaxies of the Milky Way are vertically orbiting with $f_{\rm Sat, vertical} = 0.87$ and tend to align in a plane with high azimuthal anisotropy $\xi_{\rm Sat, \phi} = 0.55$. 

\begin{figure*}
\centering\includegraphics[width=13cm]{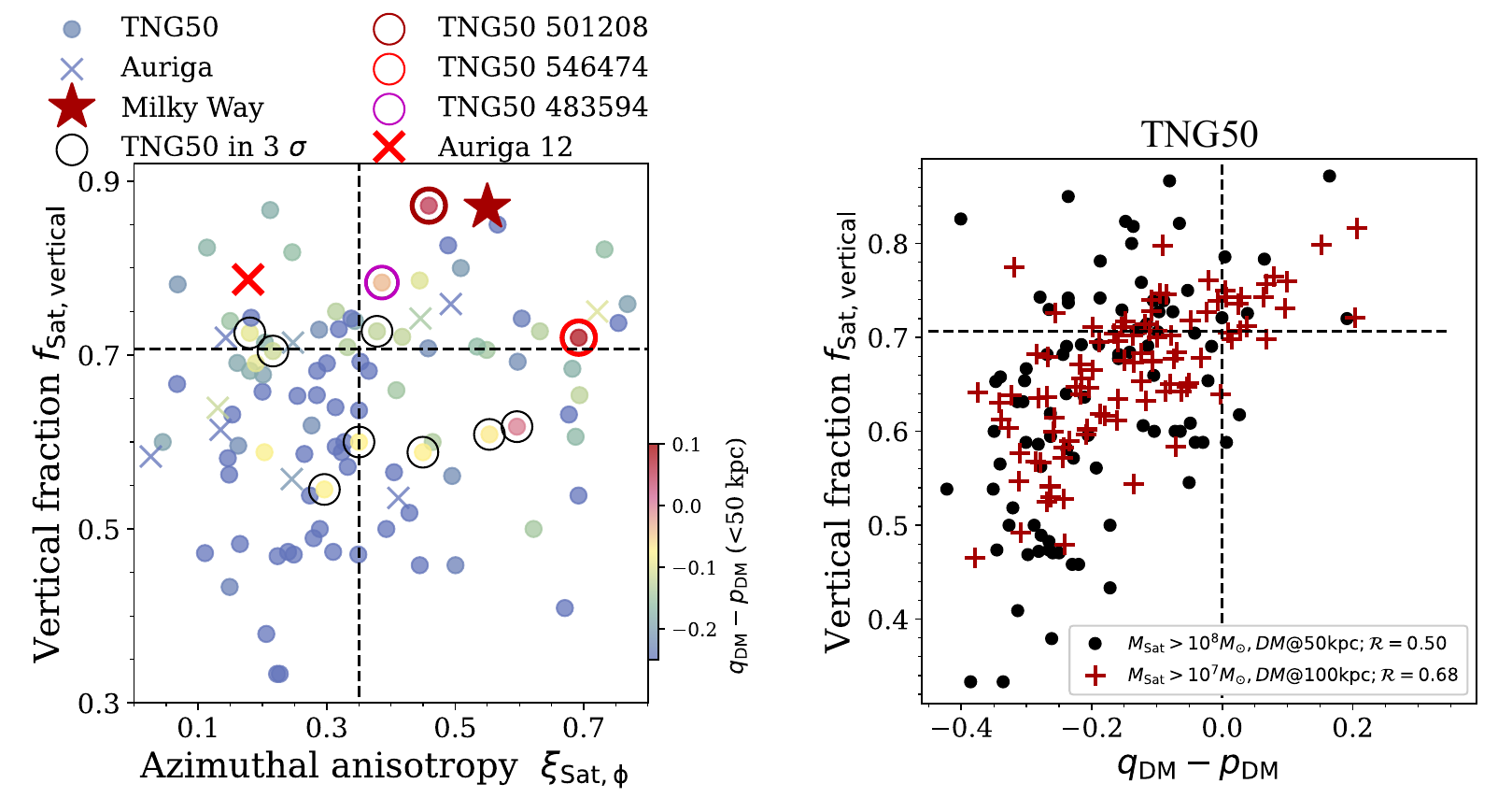}
\caption{\textbf{Arrangement of satellite system in the Milky Way and the Milky Way-like simulated galaxies, and relation with the 3D shape of DM halo.} {\bf Left:} the configuration of satellite system of each galaxy is characterised by two parameters: the fraction of satellites moving on vertical orbits $f_{\rm Sats, vertical}$, and the azimuthal anisotropy of the satellite distribution $\xi_{\rm Sat, \phi}$. Each dot is one galaxy coloured by their DM halo shape $q_{\rm DM} - p_{\rm DM}$ ($<50$ kpc). The red star represents the Milky Way. The 13 TNG50/Auriga galaxies having DM halo shapes within $3\sigma$ confidence level of the Milky Way are highlighted. {\bf Right:} correlation of the DM halo shape $q_{\rm DM} - p_{\rm DM}$ versus satellite distribution $f_{\rm Sats, vertical}$ of the 94 TNG50 galaxies, $\mathcal{R}$ in the legend is the Pearson correlation coefficiant. The correlation becomes stronger when increasing the number of satellites by taking a lower mass threshold and measuring DM shape at a larger radius.  }
\label{fig:pq_sat_ana}
\end{figure*}

 The results of these 94 TNG50 and 11 Auriga galaxies are presented in Fig.\ref{fig:pq_sat_ana}.  
As shown in the figure, the distribution of satellite galaxies around most galaxies is not entirely isotropic.
  Note that these parameters are affected by the number of satellites and there are still certain scatters in describing the satellite arrangement. However, we notice that there is a general correlation between the shape of the DM halo and the arrangement of the satellite system; the correlation is stronger for the DM shape measured at a larger radius. 
 In galaxies with oblate DM halos ($q_{\rm DM} < p_{\rm DM}$), satellite galaxies tend to align coplanarly with the stellar disc, characterised by $f_{\rm Sat, vertical}<0.707$ and $\xi_{\rm Sat, \phi} \lesssim 0.3$. This configuration is typical for systems with non-zero total angular momentum, where the orientation of the stellar disc, DM halo, and satellite system are horizontally synchronised, consistent with previous results \citep{Shao2016MNRAS.460.3772S}. There are a number of galaxies with satellites distributed highly coplanar with $f_{\rm Sat, vertical}<0.5$; a plane of satellites like the Milky Way (not considering its orientation) is thus not so rare, as also revealed by previous studies \citep{Sawala2023NatAs...7..481S}. 
However, galaxies with triaxial or near-spherical DM halos ($q_{\rm DM} \simeq p_{\rm DM}$) show a more isotropic distribution of satellites ($f_{\rm Sat, vertical}\sim 0.707$). Meanwhile, in the rare cases of galaxies with DM halo vertically aligned with the stellar disc ($q_{\rm DM}>p_{\rm DM}$), their satellite systems appear vertically orientated $f_{\rm Sat, vertical}>0.707$, and at the same time show considerable anisotropy in the azimuthal direction (large value of $\xi_{\rm Sat, \phi}$). 

\begin{figure*}
\centering\includegraphics[width=18cm]{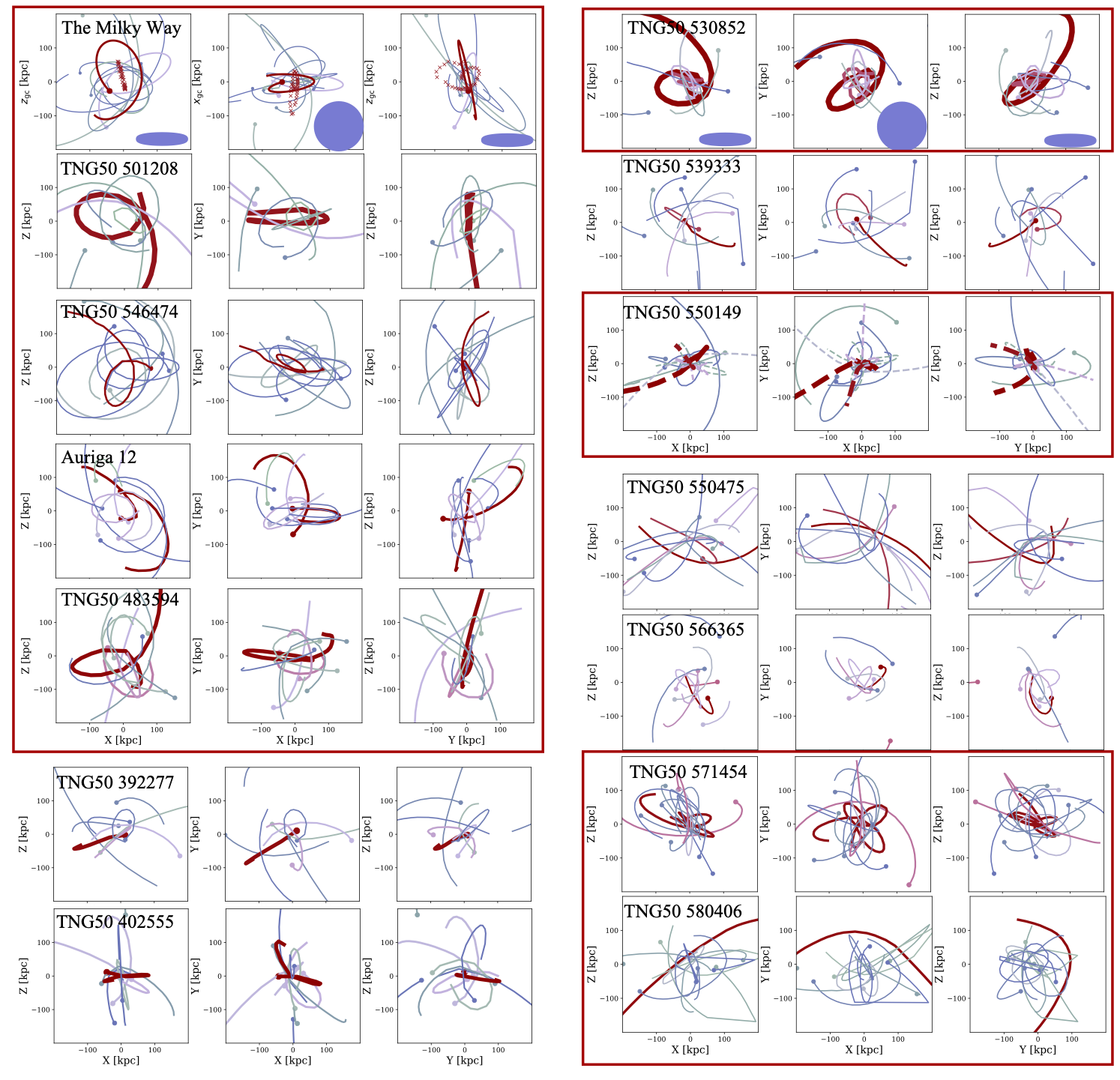}
\caption{Orbital configuration of satellites in the Milky Way and TNG50/Auriga galaxies within the $3\sigma$ confidence level. For each galaxy, we show the orbits of the ten brightest/most massive satellites. For the Milky Way, the pluses represent the orbit of the Sagittarius stream. The colours represent the relative mass of the satellites in each galaxy, red is the most massive one; the thickness of the curves are scaled with the maximum history mass of the satellites. The blue ellipses in the top panels indicate the orientation of stellar disc. The first four simulated galaxies in the top-left panels are the four galaxies we discussed, they show similar vertical satellite plane as the the Milky Way. While the rest nine simulated galaxies show rather isotropic distribution of the satellite galaxies. Eight of the 13 galaxies (highlighted by red boxes), experienced disc tilt from $30\degree$ to $100\degree$, most of them are likely induced by a massive minor merger (red curves), mostly spiraled inward on tangential orbits, except 580406.
}
\label{fig:3sigma}
\end{figure*}

For Milky Way and the 13 TNG50/Auriga galaxies within the $3\sigma$ confidence level, this analysis of the overall satellite populations with $f_{\rm Sat, vertical}$ and $\xi_{\rm Sat, \phi}$ is generally consistent with the orbital configurations of the ten brightest/most massive satellites (see Fig.\ref{fig:3sigma}). The Milky Way, TNG50 501208, TNG50 546474, and TNG50 483594 show clear planes in the orbits of their ten most massive satellites, while their overall satellite populations are also vertically aligned and highly anisotropic. Auriga 12 also exhibits a plane formed by the ten most massive satellites; its overall satellite populations are also vertically aligned, whereas the azimuthal anisotropy is not so significant. Among these, TNG50 501208 and TNG50 546474 are the closest to the Milky Way when considering either the arrangement of the ten most massive satellites or the overall satellite populations.
For the nine remaining galaxies located in $3\sigma$, there are no clear vertical satellite planes formed by the ten most massive satellites, and their overall satellite populations are generally distributed rather isotropically or coplanar aligned, as revealed by $f_{\rm Sat, vertical}$ and $\xi_{\rm Sat, \phi}$.

\section{The other TNG50 and Auriga galaxies located within $3\sigma$ confidence level}
\label{ap:3sigma}

Despite the four galaxies with DM halos and satellite planes vertically aligned due to disc tilt, there are another nine galaxies, TNG50 392277, 402555, 530852, 539333, 550149, 550475, 566365, 571454, and 580406 with DM shape outside of the $1\sigma$ but within the $3\sigma$ confidence level of the Milky Way. These galaxies have intrinsically near-spherical DM halos. Four of them, TNG50 530852, 550149, 571454, 580406 also experienced significant disc tilt with tilt angles from $30\degree$ to $100\degree$. The disc tilts of the former three galaxies are associated with and likely induced by a massive minor merger. Such events occurred quite often and were not necessarily associated with systems exhibiting satellite planes. The disc of TNG50 580406 was tilted about $30\degree$ and was not associated with any significant mergers.

However, all nine of these galaxies have satellites near isotropically, as shown in the right panels of Figure~\ref{fig:3sigma}, distinguished from the vertical satellite plane as observed in the Milky Way. This is consistent with the correlation between the DM shape and satellite arrangement we showed in the right panel of Figure~\ref{fig:pq_sat_ana}; satellite galaxies in near-spherical halos tend to be isotropically distributed.

Such galaxies with a near-spherical DM halo, although lying within $3\sigma$ confidence of our model for the Milky Way, are not likely the case for the Milky Way considering their satellite arrangements.

\end{appendix}

\end{document}